# Geometry-Resolved Atomic Oxygen Risk Assessment for Very Low Earth Orbit Spacecraft

**Authors:** Gun Hi Won[a], Hyun Jung Kim[a]*, SongYi Park[b], ChangWon Seo[b], Eunji Lee[b], SeongSik Yoon[b]

[a]Department of Aerospace Engineering, Korea Advanced Institute of Science and Technology

[b]Hanwha Systems Co., Ltd.

* corresponding author: hyunjung.kim1@kaist.ac.kr

**Abstract** Atomic oxygen (AO) is a major durability concern for spacecraft in very low Earth orbit (VLEO), yet orbit-averaged fluence does not resolve exposure on individual surfaces and internal components. This study develops a geometry-resolved AO assessment by coupling NRLMSISE-00, HWM07, and SYSTEMA ATOMOX. One-year simulations were performed for a 350 km circular Sun-synchronous orbit at LTAN 06:00 and 12:00 using a baseline spacecraft, a wedge-modified body, and two synthetic aperture radar antenna sub-arrays. The LTAN 12:00 orbit produced 8–10% higher orbit-averaged AO flux than LTAN 06:00. For the baseline geometry, the ram-facing surface accumulated 6.9–7.5 × $10^{21}$ atoms/ $cm^2$, whereas side and zenith/nadir surfaces received only 3–5% of the ram fluence. Material-specific erosion yields changed the component-level risk ranking: the CFRP zenith panel was predicted to erode by 15.1–16.2 μm/ year despite receiving much lower fluence than the ram-facing multilayer insulation. The wedge generated approximately one order of magnitude spatial variation through local shielding. Housing openings also allowed AO to reach internal printed circuit boards, with maximum annual fluences of 9.5 × $10^{16}$ and 4.0 × $10^{19}$ atoms $cm^{-2}$ in the H- and V-polarized antenna models, respectively. HWM07 winds produced 10–20% side-panel asymmetry, which decreased below 1% when winds were disabled. Comparison with MISSE-8 reproduced the measured zenith-to-ram ratio of approximately 4% but underpredicted wake exposure, identifying a limitation of ballistic ray tracing. These results demonstrate that VLEO AO durability requires coupled consideration of orbit, atmospheric winds, geometry, and material response.

## 1. Introduction

The paradigm of the global space industry is rapidly shifting from large, monolithic satellites to cost-effective, small satellite constellations. This transition is largely driven by the New Space movement, which emphasizes rapid development cycles and reduced launch costs. In this context, Very Low Earth Orbit (VLEO), typically defined as altitudes below 450 km, has emerged as a highly attractive operational domain [1]. Operating in VLEO provides substantial benefits for Earth observation and telecommunication missions. For optical and radar payloads, the reduced distance to the target significantly enhances spatial resolution and signal-to-noise ratio (SNR), allowing for the use of smaller, less powerful instruments [2]. Furthermore, communication satellites benefit from lower latency and improved link budgets.

Despite these advantages, the VLEO environment poses formidable engineering challenges. Unlike traditional Low Earth Orbit (LEO) where radiation is the primary concern, VLEO is characterized by a relatively dense residual atmosphere. This environment induces significant aerodynamic drag, necessitating continuous orbit maintenance, and exposes the spacecraft to high concentrations of atomic oxygen (AO) [3]. AO is formed by the photodissociation of diatomic oxygen by solar ultraviolet radiation and constitutes the dominant atmospheric species in the VLEO region [4]. Figure 1 illustrates the variations in atmospheric constituent density, AO flux with altitude, and the effects of solar activity on AO density and flux.

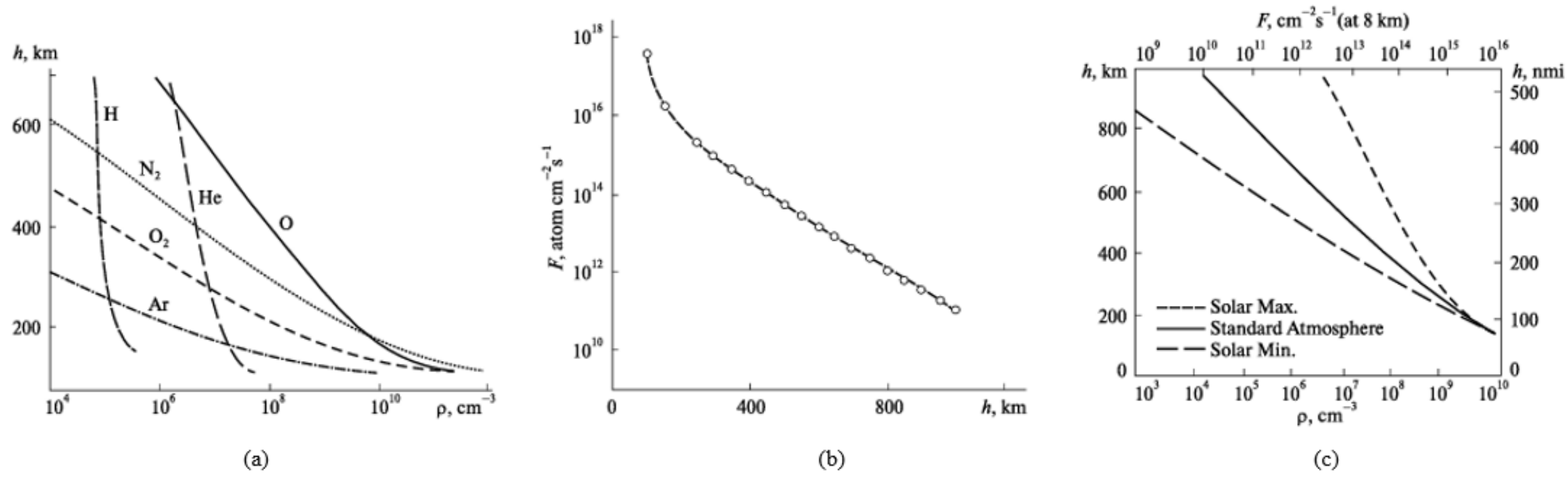

*Figure 1. (a) Variation of density* ρ *of atmospheric constituents with altitude h, (b)Variation of AO flux F with altitude h, (c) Variation of AO density* ρ *and flux F with solar activity [5]*

Spacecraft in VLEO travel at orbital velocities of approximately 7.8 km/s. Consequently, the ambient AO impinges on the ram-facing surfaces with a relative kinetic energy of approximately 4 to 5 eV [6]. This hyperthermal AO is highly reactive and causes severe degradation of various spacecraft materials, particularly polymers, carbon-fiber-reinforced polymers (CFRP), and certain optical coatings. The degradation mechanisms include chemical oxidation, which leads to mass loss and the formation of volatile products (e.g., CO, $CO_2$, $H_2O$), and physical erosion, which increases surface roughness [4]. This surface roughening, often characterized by a "carpet-like" morphology, alters the thermo-optical properties (solar absorptance and thermal emittance) of the materials, potentially compromising the spacecraft's thermal control system [7].

Several international missions have highlighted the critical impact of AO in VLEO. The European Space Agency’s (ESA) Gravity Field and Steady-State Ocean Circulation Explorer (GOCE) operated below 270 km altitude using an aerodynamic spacecraft design and ion propulsion-based drag-free control to compensate for atmospheric drag. [8] [9]. The Japan Aerospace Exploration Agency's (JAXA) Super Low Altitude Test Satellite (SLATS), also known as "Tsubame," successfully operated at altitudes down to 167.4 km. SLATS carried the Atomic Oxygen Monitor (AMO), which, included the Atomic Oxygen Fluence Sensor (AOFS) and the Material Degradation Monitor (MDM) to measure the AO environment and observe the in-orbit degradation of candidate materials for future super-low-altitude missions. These data provide valuable knowledge on AO fluence and its effects on spacecraft materials, supporting material selection for future VLEO satellite design [10] [11]. Furthermore, long-term observations from the International Space Station (ISS), orbiting at approximately 400 km, have continuously documented the erosive effects of AO on solar arrays and external insulation [12]. Table 1 provides a brief summary of representative VLEO satellite missions.

| Satellite | Operator | Orbit Information | Operational Status | Size | Major Mission & Achievements | Ref. |
| --- | --- | --- | --- | --- | --- | --- |
| GOCE<br>(Gravity field and steady-state Ocean Circulation Explorer) | European Space Agency (ESA) | Below 270 km;<br>Circular Sun-Synchronous dawn-dusk orbit;<br>Inclination 96.7° | Mission completed;<br>Launched Mar. 2009 | 5.3m long;<br>Cross-section $1.1m^2$ | - Gravity anomaly accuracy 1mGal<br>- Spatial resolution ≤100 km<br>- Aerodynamic configuration and ion-thruster drag-free control | [8]<br>[9] |
| SLATS<br>(Super Low Altitude Test Satellite)<br>(TSUBAME) | Japan Aerospace Exploration Agency (JAXA) | ~167.4 - 271.7 km altitude;<br>Maintained 167.4 km for 7 days | Mission completed;<br>Launched Dec. 23 2017;<br>Ended Oct. 1, 2019 | 2.5m x 5.2m x 0.9m;<br>400 kg | - Demonstrated VLEO operation<br>- Ion-engine drag compensation<br>- Atmospheric-density measurement | [10]<br>[11] |
| DANDE<br>(Drag and Atmospheric Neutral Density Explorer) | University of Colorado | Target altitude 200 - 350 km; 350km circular Sun-synchronous noon-to-midnight orbit | Mission completed | Nanosat;<br>~50 kg | - In-situ measurement of thermospheric neutral density, wind, and atmospheric composition | [13] |
| Aeolus | European Space Agency (ESA) | 320 km altitude;<br>Sun-Synchronous dusk-dawn orbit | Mission completed;<br>Launched Aug. 22nd 2018 | ~1,200 kg | - First satellite mission to measure global wind profiles using spaceborne Doppler wind lidar (Aladin) | [14] |
| ISS<br>(Internation Space Station) | National Aeronautics and Space Administration (NASA) | ~400km altitude | Multiple missions completed / ongoing | ~420,000 kg | - Space material degradation and erosion database development | [15] |
| SOAR<br>(Satellite for Orbital Aerodynamics Research) | University of Manchester | Initial orbit 421-415km orbit; Naturally decayed to 226-234km before re-entry | Mission completed;<br>Launched June 3 2021;<br>Ended Mar. 14 2022 | 3U Cubesat | - Aerodynamic material characterization in VLEO<br>- Drag/lift coefficient measurements using steerable fins | [16]<br>[17]<br>[18] |
| CHAMP<br>(Challenging Minisatellite Payload) | German Aerospace Center (DLR) | Initial altitude ~454km;<br>Decayed to below 300km;<br>Inclination 87.3° | Mission completed;<br>Launched July 15, 2000;<br>Re-entered Sep. 19, 2010 | ~522kg;<br>~4.3m body with 4m deployed boom | - Gravity field, magnetic field, atmosphere and ionosphere observations | [19]<br>[20] |

*Table 1. Summary of representative VLEO satellite missions.*

To ensure the survivability and performance of VLEO spacecraft, it is imperative to accurately predict the AO exposure levels and the resulting material degradation. While ground-based AO simulation facilities are essential for material testing, they often struggle to perfectly replicate the complex orbital environment, including the exact energy distribution and angular incidence of the AO flux [5] [21]. Therefore, numerical simulations that account for the specific orbital parameters and the complex geometry of the spacecraft are crucial. The geometry of the spacecraft dictates the local incidence angle of the AO flux and introduces shielding effects, leading to a highly non-uniform distribution of AO fluence across the spacecraft surfaces [22].

Although extensive AO studies have characterized material erosion in LEO and VLEO, most published assessments report orbit-averaged fluence or material-specific erosion values, often derived from flat-panel exposure experiments such as MISSE or laboratory plasma sources. These approaches provide valuable guidance for material selection, but they rarely capture the local variations in AO flux induced by complex spacecraft geometry. For VLEO platforms, structural appendages, antenna housing, and openings can redistribute local AO flux and cumulative fluence, producing highly non-uniform surface exposure that is difficult to reproduce in ground-based AO simulation facilities. This motivates the need for numerical, geometry-resolved AO simulations that explicitly account for spacecraft configuration, attitude, and orbital parameters when evaluating surface erosion and durability.

Therefore, this study aims to quantitatively assess AO exposure and material erosion risk as a function of spacecraft geometry and Local Time of Ascending Node (LTAN), establishing a geometry-resolved AO risk assessment framework to inform VLEO spacecraft design and material selection. Utilizing the NRLMSISE-00 atmospheric, HWM07 wind model and the SYSTEMA ATOMOX simulation suite, we evaluate the AO flux and cumulative fluence on various spacecraft configurations, including a baseline shape, a modified shape with structural appendages, and detailed Synthetic Aperture Radar (SAR) antenna sub-arrays. The analysis considers a 350 km circular Sun-Synchronous Orbit (SSO) and investigates the influence of different Local Time of Ascending Node (LTAN) conditions. LTAN values of 06:00 and 12:00 were selected to represent two commonly adopted operational scenarios for VLEO missions, corresponding to dawn-dusk SAR missions and near-noon Earth observation missions, respectively. The findings provide critical insights for the geometric design and material selection of future VLEO satellites.

The main contribution of this study is to establish a geometry-resolved AO durability assessment approach for VLEO spacecraft at the design stage. Unlike orbit-averaged AO fluence estimates or material-only erosion assessment, the analysis resolves local AO exposure over realistic spacecraft surfaces and links the resulting fluence distribution to material-dependent erosion risk. The study provides three specific contributions. First, it quantifies how LTAN-dependent AO environments modify surface level fluence and erosion risk for a representative VLEO spacecraft. Second, it demonstrates how structural appendages and antenna geometries redistribute AO exposure through shielding, shadowing, and cavity-penetration effects. Third, it identifies wind-induced side-panel exposure asymmetry using HWM07 on/off comparisons, clarifying the role of atmospheric winds in local AO incidence. These contributions provide design-relevant guidance for material selection, protective-coating placement, and payload configuration of VLEO spacecraft.

## 2. Methodology

2.1. Atomic Oxygen Environment Modeling and Analysis

The quantitative assessment of AO-induced degradation requires an accurate estimation of the AO flux and fluence. The AO flux ($\phi_{AO}$), defined as the number of oxygen atoms impacting a unit area per unit time (atoms/cm²/s), is calculated as the product of the local AO number density ($n_{AO}$) and the relative velocity of the spacecraft ($v_{AO}$) [23].

$$\phi_{AO} = n_{AO} \cdot v_{AO} \quad (1)$$

The cumulative AO fluence ($\psi_{AO}$), representing the total number of atoms impacting a unit area over a specific mission duration (atoms/cm²), is obtained by integrating the flux over time. The actual erosion depth of a material can then be estimated by multiplying the fluence by the material's specific erosion yield ($E_y$, cm³/atom).

To model the AO density in the VLEO environment, the NRLMSISE-00 empirical atmospheric model was employed by MATLAB. NRLMSISE-00 is widely recognized for its accuracy in predicting atmospheric composition, temperature, and density from the ground to the exosphere, taking into account variations in solar activity, geomagnetic activity, latitude, longitude, and time [24]. For this study, the environmental conditions were set as nominal to represent a moderate solar activity level, with a solar radio flux index ($F10.7$) of 150 and a planetary geomagnetic index ($A_p$) of 15. The reference epoch for the simulations was set to January 1, 2028 considering it as a near-term date with potential launch feasibility. These conditions were selected to represent atomic oxygen density and flux levels generally expected in the VLEO environment, with the objective of analyzing atomic oxygen effects under representative operational conditions rather than under extreme-case scenarios.

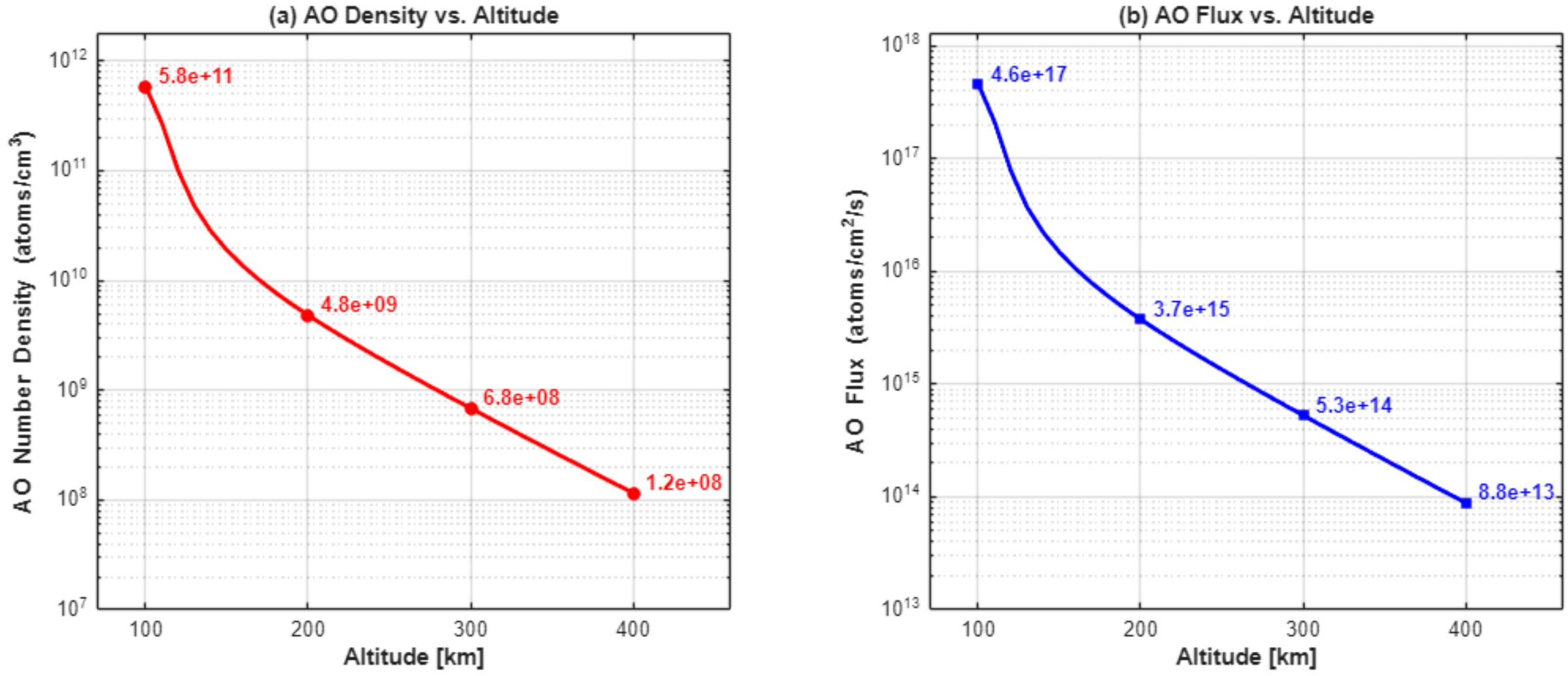

*Figure 2. AO number density and flux as a function of altitude, calculated using the NRLMSISE-00 model (Epoch: 2028.01.01, F10.7=150, Ap=15, longitude=127, latitude=37).*

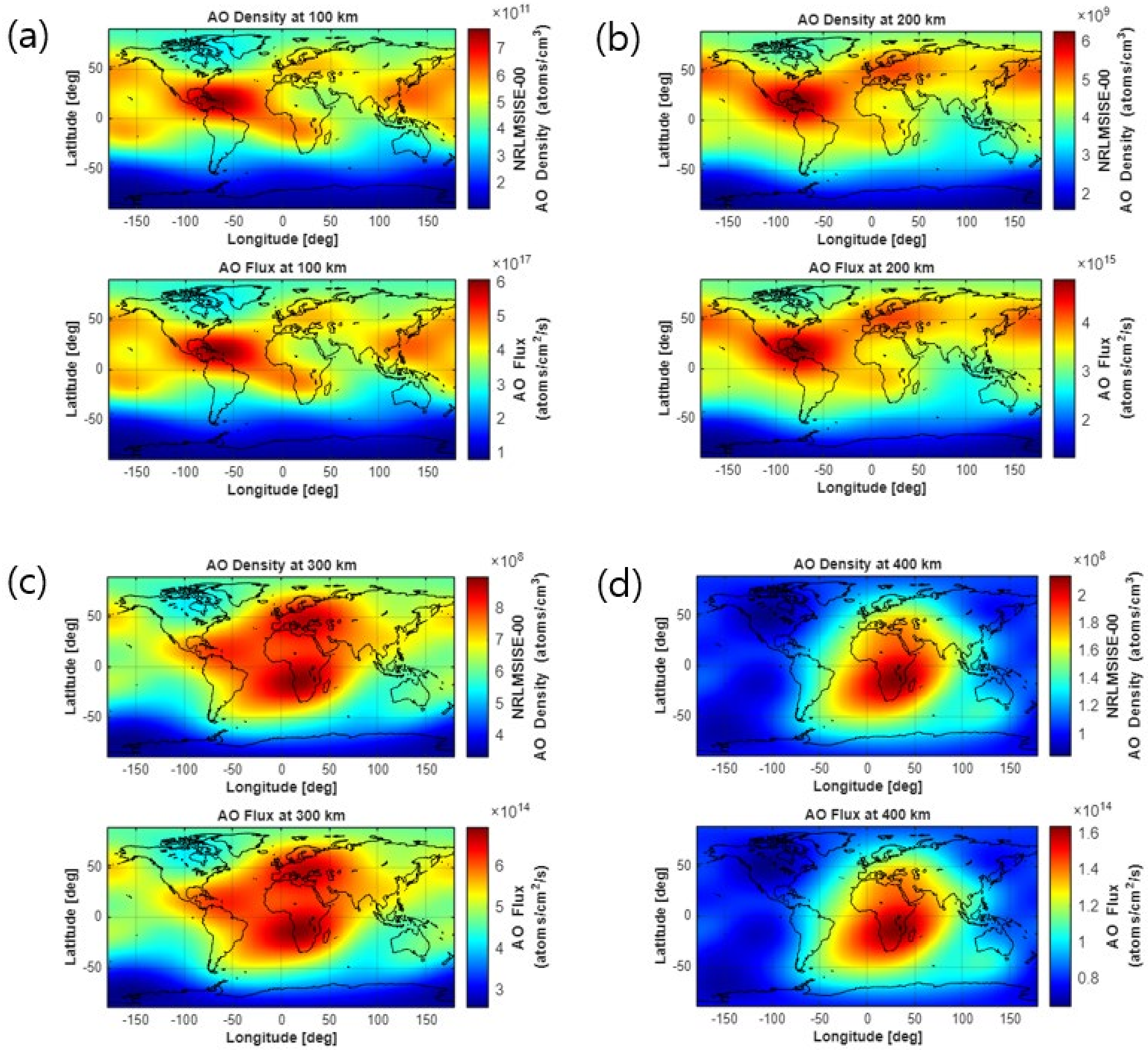

*Figure 3. Altitudinal distributions of AO density (top) and flux (bottom) on January 1, 2028, UT 12:00 based on the NRLMSISE-00 model at (a) 100km, (b) 200km, (c) 300km, and (d) 400km.*

Figure 2 presents the atomic oxygen number density and the corresponding atomic oxygen flux calculated using the orbital velocity of a spacecraft in a circular orbit at each altitude. Figure 3 shows the spatial distributions of atomic oxygen density and flux as functions of latitude and longitude at selected altitudes under the prescribed environmental conditions. Relatively enhanced AO density and flux are observed in specific latitude-longitude regions, reflecting the combined effects of thermospheric composition, temperature structure, and local-time-dependent atmospheric variation The averaged density and flux values are summarized in Table 2, confirming that atomic oxygen density and flux increases as the orbital altitude decreases. At an altitude of 300 km, the average atomic oxygen density was approximately $7.2 \times 10^8\ atoms/cm^3$, while the average atomic oxygen flux was approximately $5.6 \times 10^{14}\ atoms/cm^2/s$. Figure 4 shows the altitude dependence of the average equatorial AO density and flux calculated using the NRLMSISE-00 atmospheric model.

| Altitude | 100km | 200km | 300km | 400km |
|---|---|---|---|---|
| Average AO density $(atoms/cm^3)$ | $5.4 \times 10^{11}$ | $4.5 \times 10^{9}$ | $7.1 \times 10^{8}$ | $1.4 \times 10^{8}$ |
| Average AO flux $(atoms/cm^2/s)$ | $4.3 \times 10^{17}$ | $3.5 \times 10^{15}$ | $5.5 \times 10^{14}$ | $1.1 \times 10^{14}$ |

*Table 2. Average Equatorial Atomic Oxygen Density and Flux at Different Altitudes*

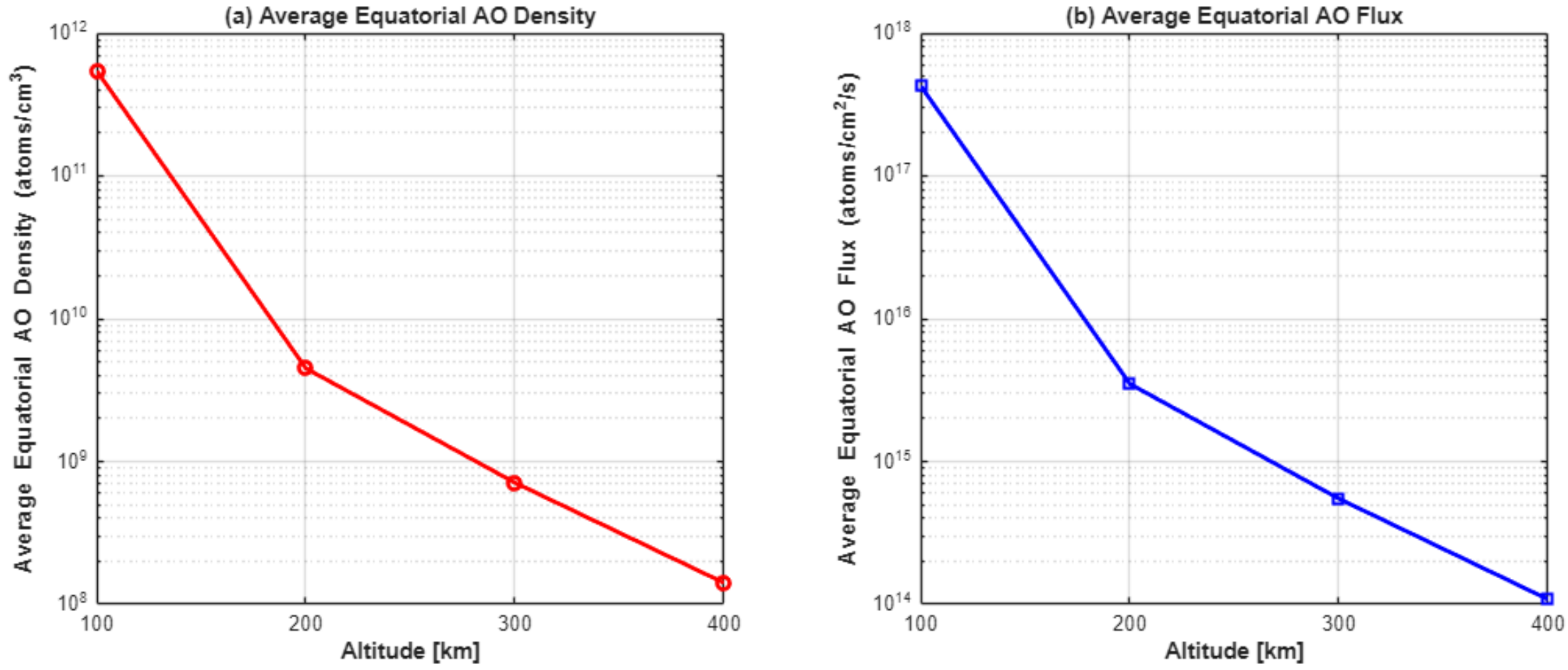

*Figure 4. Altitude dependence of average equatorial AO density and flux calculated using the NRLMSISE-00 atmospheric model (Epoch: 2028.01.01, F10.7=150, Ap=15)*

Table 3 summarizes the atomic oxygen environment analysis based on the satellite orbital conditions used in the subsequent analyses. The target orbit was a 350 km circular Sun-synchronous orbit (SSO) which is commonly used for VLEO satellite. To investigate the effect of the local time of the ascending node (LTAN), LTAN values of 6:00 and 12:00 were separately applied, enabling comparison of the atomic oxygen distribution under targeted orbit (altitude 350km, circular orbit, sun-synchronous, 2028.01.01 00:00 UTC as starting datetime) conditions but different solar illumination environments. The reported values represent orbit-averaged atomic oxygen density and flux over a one-year simulation period starting from the reference epoch. The orbital propagation was performed using a fixed time step of 60 seconds.

| Altitude | LTAN | Average AO density $(atoms/cm^3)$ | Average AO flux $(atoms/cm^2/s)$ |
|---|---|---|---|
| 350 km (Sun-Synchronous Orbit) | 6:00 | $2.8 \times 10^{8}$ | $2.2 \times 10^{14}$ |
| | 12:00 | $3.1 \times 10^{8}$ | $2.4 \times 10^{14}$ |

*Table 3. One-year orbit-averaged AO density and flux for a 350 km Sun-synchronous orbit*

Initial analysis using the NRLMSISE-00 model confirmed the exponential increase in AO density with decreasing altitude (Figure 2). At the target altitude of 350 km, the average AO density was calculated to be approximately $2.8 \times 10^{8}$ atoms/cm³ for LTAN 06:00 and $3.1 \times 10^{8}$ atoms/cm³ for LTAN 12:00. The corresponding average fluxes were $2.2 \times 10^{14}$ atoms/cm²/s and $2.4 \times 10^{14}$ atoms/cm²/s, respectively. Differences in the orbit-averaged AO density and flux were observed between the LTAN conditions, indicating that the local-time configuration of the orbit can influence the baseline AO environment experienced by a spacecraft. Since LTAN determines the local solar time sampled along the orbit, the resulting AO environment reflects the combined effects of orbital geometry and local-time dependent thermospheric variations predicted by the NRLMSISE-00 model.

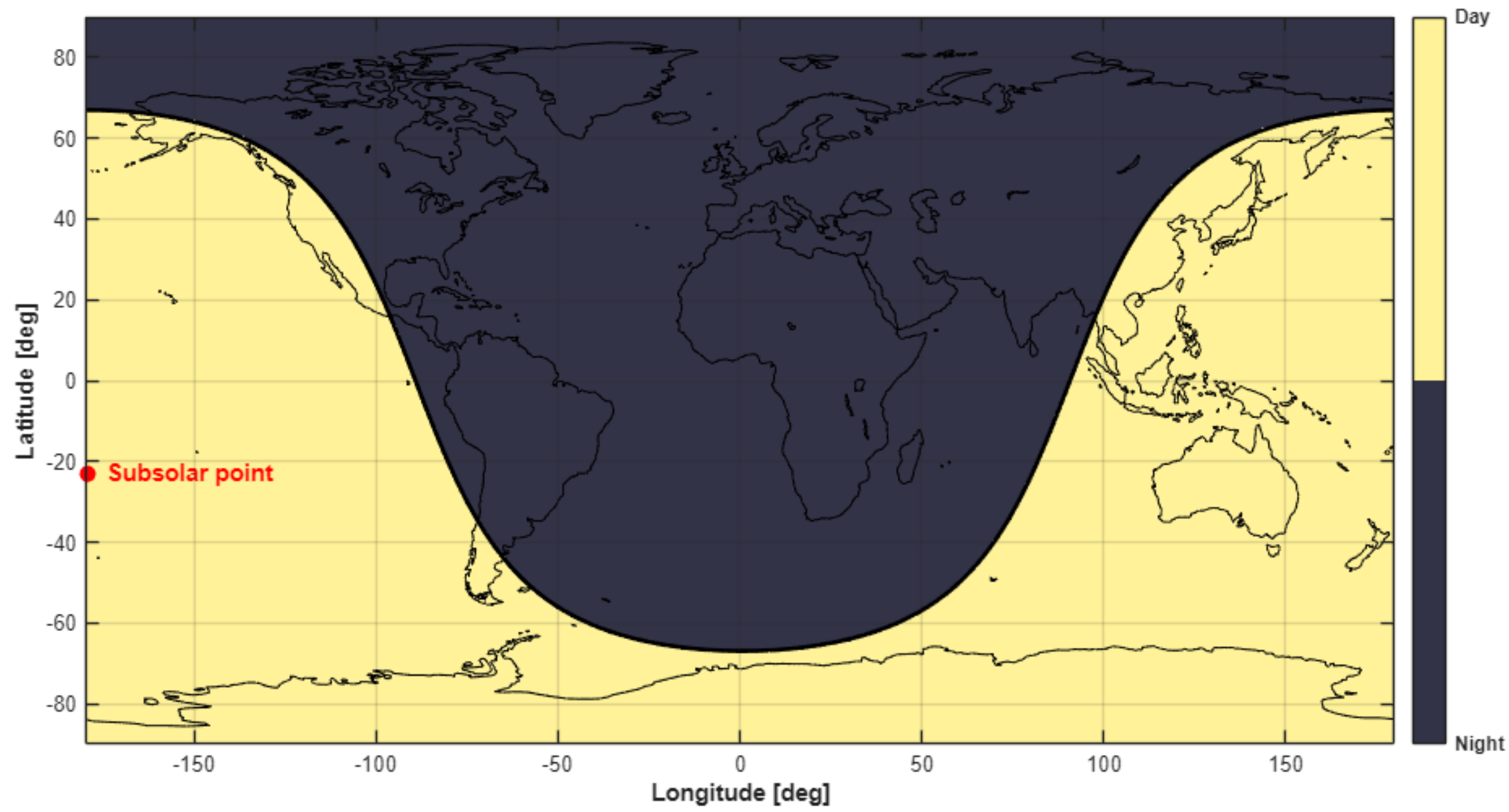

*Figure 5. Day/Night distribution on 2028.01.01 00:00 UTC*

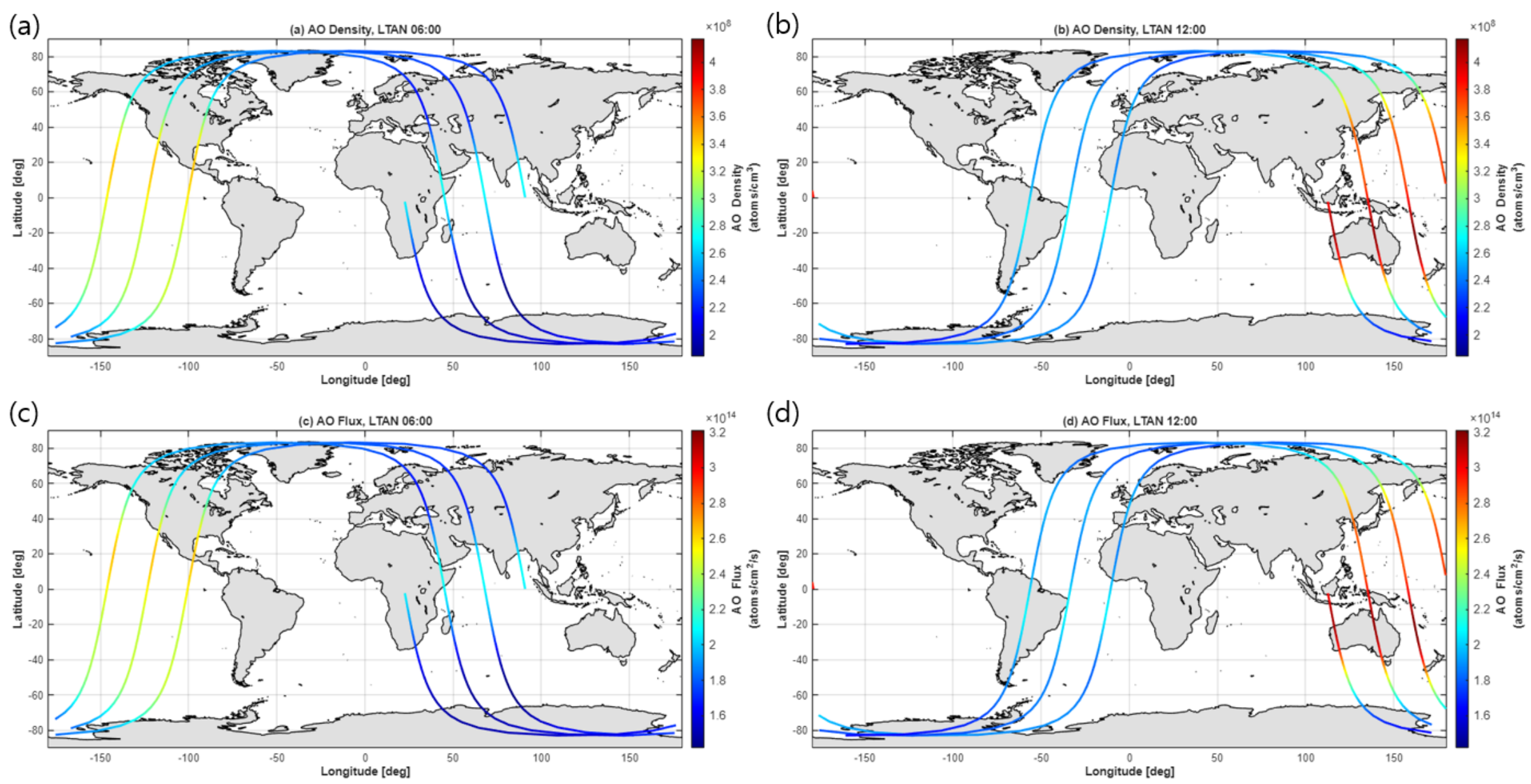

*Figure 6. Comparison of AO density and flux distributions along the satellite orbit under different LTAN conditions: (a, c) LTAN 6:00 case and (b, d) LTAN 12:00 case*

Figure 6 presents the modeled atomic oxygen environment along three orbital periods for the two Sun-synchronous orbit configurations considered in this study. Figure 6(a) and 6(b) show the atomic oxygen density distributions for the LTAN 06:00 and LTAN 12:00 cases, respectively, while Figure 6(c) and 6(d) present the corresponding atomic oxygen flux distributions. The colored orbital tracks indicate the sampled spacecraft positions, with the color representing the local AO density of flux predicted by the NRLMSISE-00 model.

Figure 5 illustrates the global day-night distribution at the reference epoch of 2028.01.01 00:00 UTC, showing the illuminated and non-illuminated regions of the Earth. When Figures 5 and 6 are considered together, it can be seen that relatively higher atomic oxygen density and flux generally occur along the sunlit portions of the orbital tracks, whereas lower values are observed in regions located on the nightside of the Earth. This behavior is physically consistent with the primary production mechanism of thermospheric atomic oxygen, which is generated through the photodissociation of molecular oxygen by solar extreme ultraviolet radiation.

## 2.2. Spacecraft Geometry and Simulation Setup

The numerical analysis of AO exposure on the spacecraft surfaces was conducted using the ATOMOX module within the SYSTEMA software suite developed by Airbus Defence and Space. SYSTEMA ATOMOX is a comprehensive 3D numerical tool designed to evaluate the flux and fluence of atmospheric species, including AO, on complex spacecraft geometries, accounting for shadowing and multiple reflections [25].

The baseline mission profile was defined as a circular Sun-Synchronous Orbit (SSO) at an altitude of 350 km. Two representative Local Time of Ascending Node (LTAN) configurations were considered in this study: LTAN 06:00 and LTAN 12:00. These values were selected to represent commonly used operational scenarios for VLEO spacecraft, including dawn-dusk radar remote sensing missions and near-noon Earth observation missions. The simulation duration was set to one year, while the event computation interval used in the ATOMOX analysis was set to 1 day. The spacecraft attitude was fixed, with the +x-axis aligned with the velocity vector (ram direction) and the -z-axis pointing towards nadir (Earth-pointing), representing a typical nadir-pointing operational attitude commonly adopted by Earth-observing and remote-sensing satellites. The baseline spacecraft geometry was modeled as a simplified rectangular prism with approximate dimensions of 2.0 m x 1.0 m x 0.5 m (length x width x height), representing a small satellite platform operating in VLEO. Figure 7 presents the spacecraft coordinate system used in this study and the direction of AO flux relative to the spacecraft surfaces.

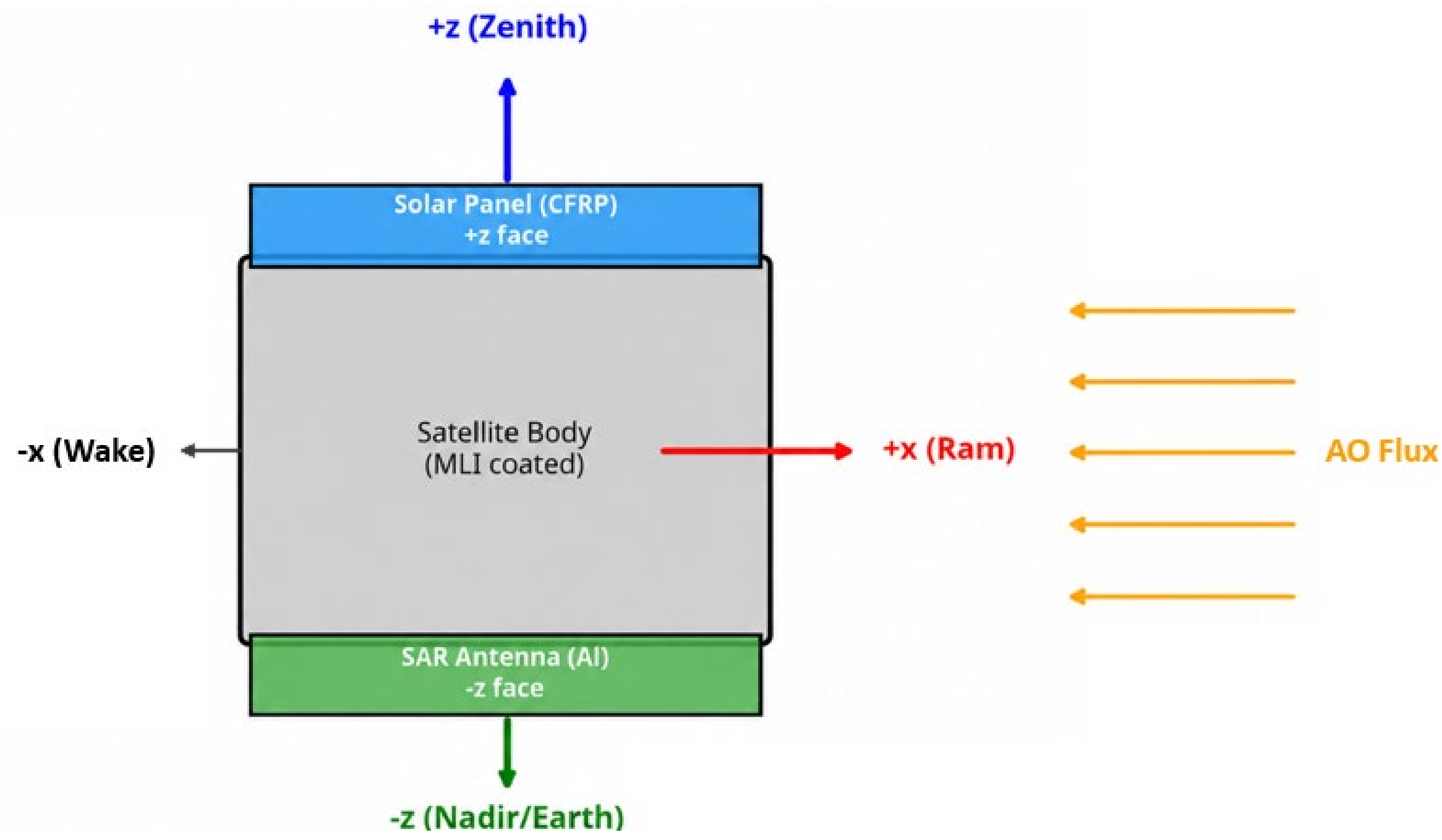

*Figure 7. Schematic of the spacecraft coordinate system and AO exposure geometry for a nadir-pointing satellite.*

Four distinct geometric cases were evaluated:

- **Case 1 (Baseline Non-deployable):** A simplified rectangular prism representing the main body of the satellite.
- **Case 2 (Wedge Addition):** The baseline shape modified with a wedge-like structure on the ram-facing (+x) surface.
- **Case 3 (H-pol Antenna Sub-array):** A detailed model of a horizontally polarized SAR antenna sub-array.
- **Case 4 (V-pol Antenna Sub-array):** A detailed model of a vertically polarized SAR antenna sub-array.

Figure 8 shows the four geometric configurations considered in the analysis, including the baseline, wedge-added, H-polarization sub-array, and V-polarization sub-array models.

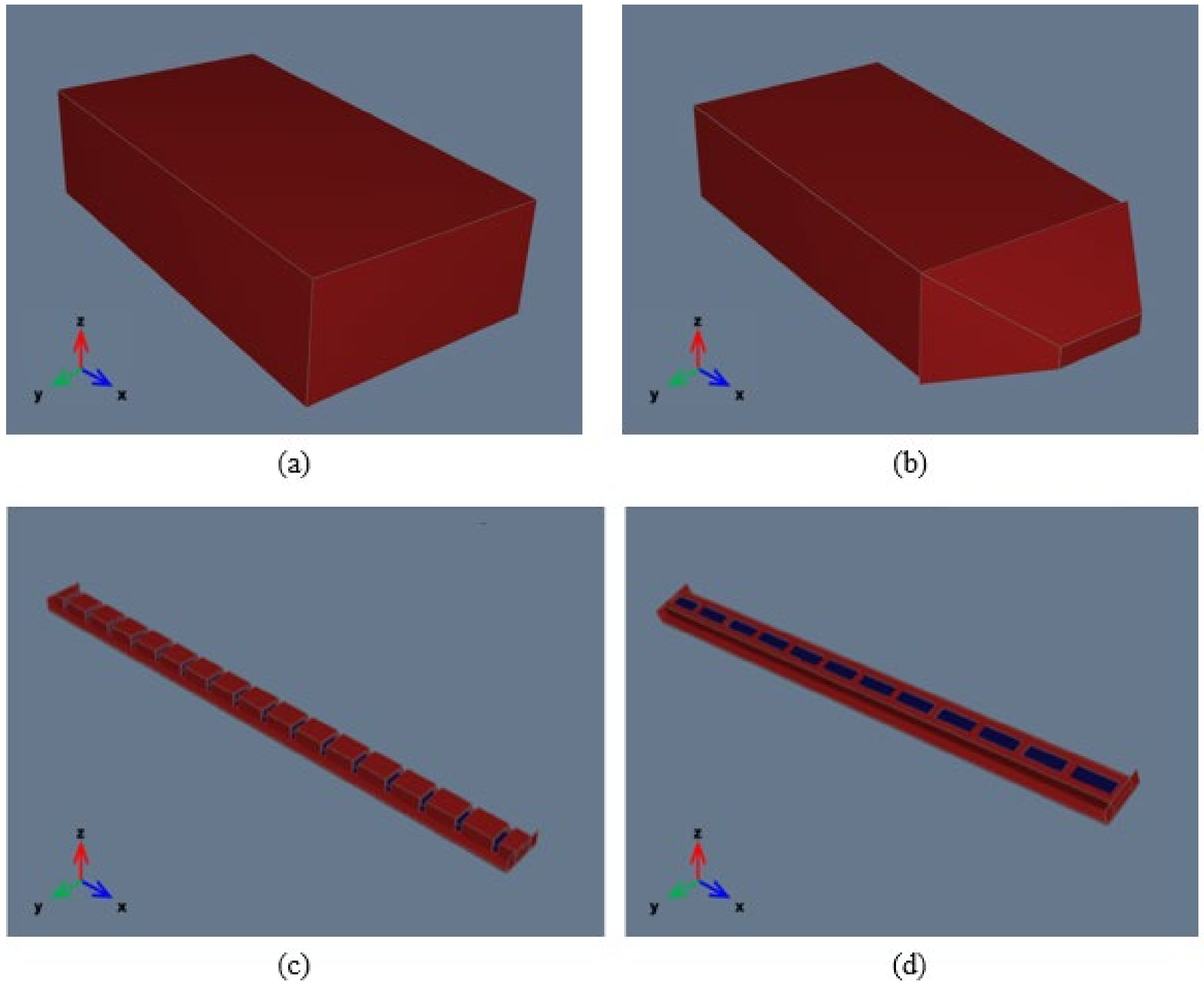

*Figure 8. Geometric models used for the analysis: (a) Case 1 (Baseline Non-deployable) (b) Case 2 (Wedge Addition) (c) Case 3 (H-pol Antenna Sub-array) (d) Case 4 (V-pol Antenna Sub-array)*

The material properties assigned to the different surfaces for the erosion calculations were based on typical spacecraft materials and their corresponding erosion yields derived from published NASA atomic oxygen erosion-yield databases and Materials International Space Station Experiment (MISSE) data [4][26]. The side panels were modeled as Multi-Layer Insulation (MLI) with a SiOx-coated aluminized polyimide outer layer ($E_y = 1.0 \times 10^{-26}$ cm³/atom). The solar arrays (+z face) were modeled as CFRP ($E_y = 5.6 \times 10^{-24}$ cm³/atom), and the antenna structures (-z face) and wedge appendages were modeled as Aluminum 6061-T6 ($E_y = 0$ cm³/atom). The solar-array was simplified as a CFRP structural panel to represent the exposed substrate/backing structure rather than the detailed solar-cell coverglass stack, since glass-based coverglass materials are generally regarded as AO-resistant and exhibit erosion yields several orders of magnitude lower than those of polymeric structural materials.

## 3. Results and Discussion

### 3.1. Effect of Baseline Geometry (Case 1)

To evaluate the influence of spacecraft geometry on AO exposure, a simplified rectangular prism model representing a non-deployable spacecraft configuration was employed as the baseline case. The +z surface was modeled as a solar array panel, the -z surface as a SAR antenna panel, and the $\pm$x and $\pm$y surfaces as MLI-covered side panels. The material properties assigned to each surface and their corresponding erosion yield values are summarized in Table 4. Because no exact material counterparts were available from the Materials International Space Station Experiment (MISSE) database, erosion yield values were selected from the most representative MISSE materials with similar compositions and expected AO response characteristics. Figure 9 shows the meshed geometry of the baseline non-deployable spacecraft model (Case 1) used for the AO exposure analysis.

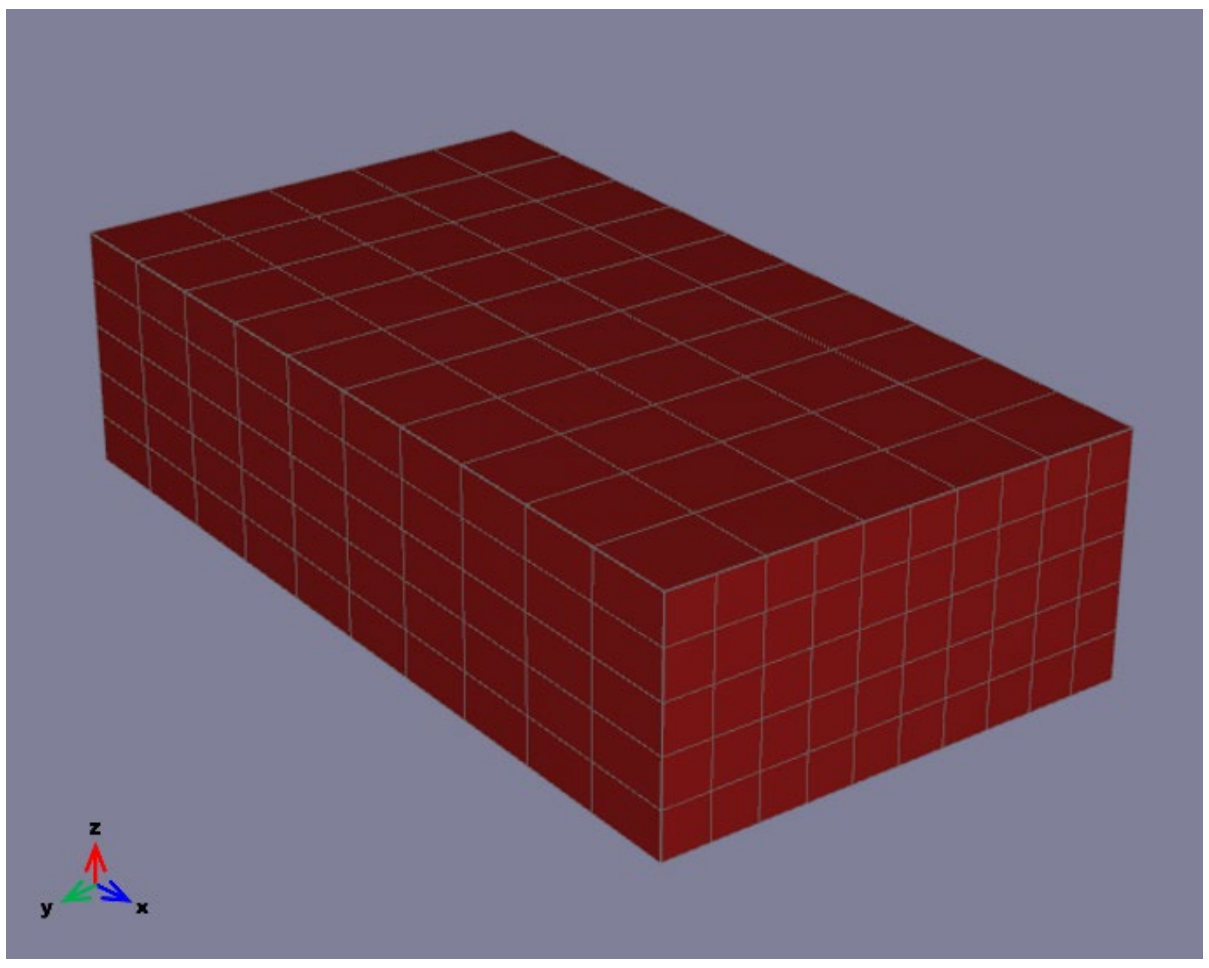

*Figure 9. Geometry of the baseline non-deployable spacecraft model (Case 1)*

| Spacecraft component | Material | Surface / Orientation | Erosion Yield $(cm^3/atom)$ | Ref. |
|---|---|---|---|---|
| Side panel | MLI | +x / Ram | $1.0 \times 10^{-26}$ | [4] |
| | | -x / Wake | | |
| | | +y / Starboard | | |
| | | -y / Port | | |
| Solar array | CFRP | +z / Zenith | $5.6 \times 10^{-24}$ | [26] |
| Antenna | Al | -z / Nadir | 0 | [4] |

*Table 4. Materials and corresponding erosion yield values assigned to the baseline non-deployable spacecraft model (Case 1)*

3.1.1. Results for the LTAN 06:00 Case

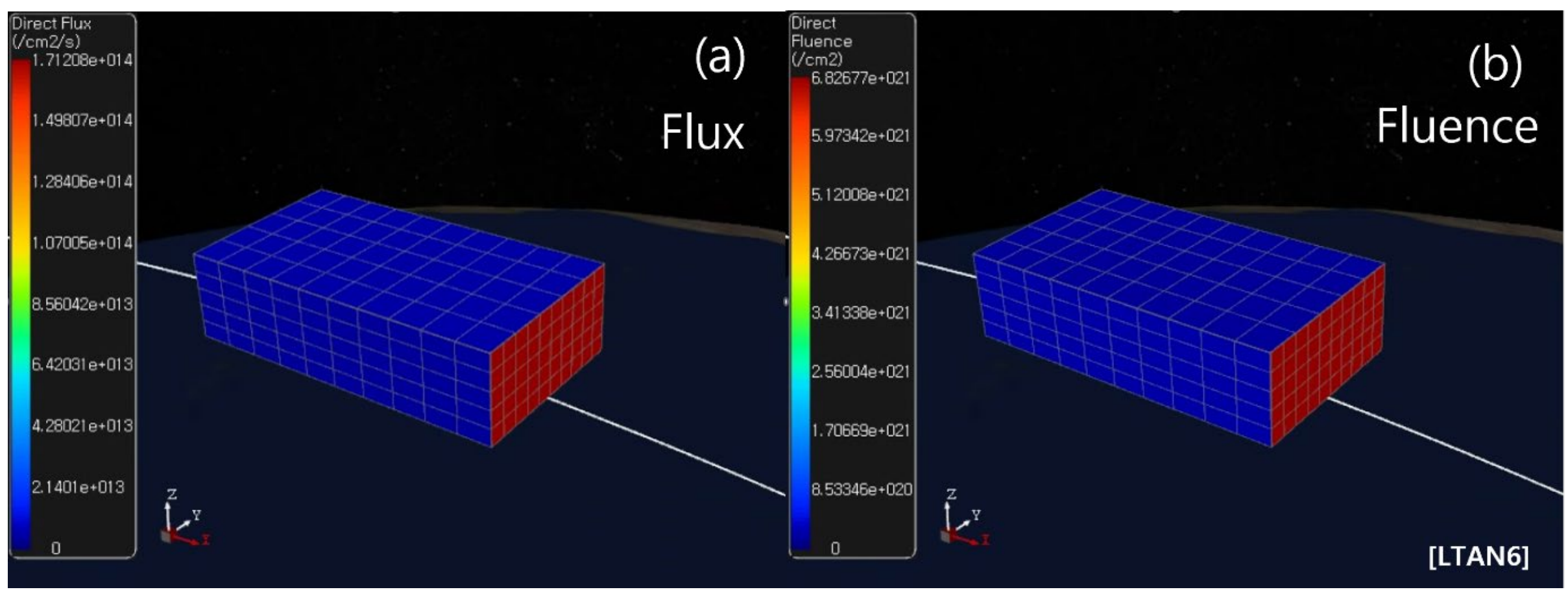

*Figure 10. Simulated AO distribution over the surface of the baseline non-deployable spacecraft model (Case 1) for the LTAN 06:00 orbit: (a) AO flux and (b) cumulative AO fluence*

The simulation results Figure 10 showed that the AO flux was nearly uniform across all mesh elements belonging to the same surface. Therefore, rather than presenting results for every mesh element, a representative mesh element was selected from each surface for quantitative analysis. The blue mesh highlighted in Figure 11 denotes the selected analysis regions and the corresponding AO flux and cumulative AO fluence values were extracted from these locations for further evaluation.

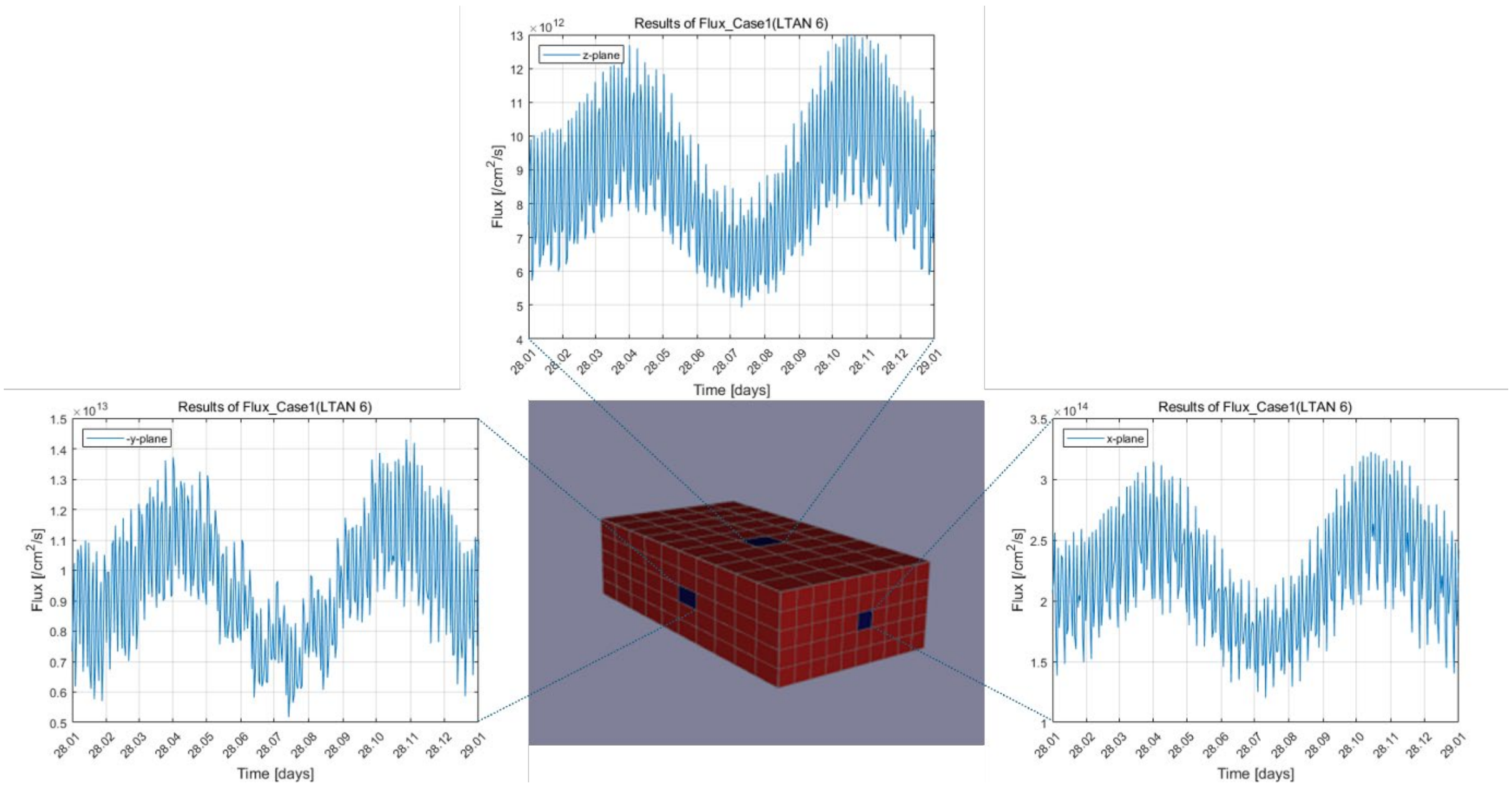

*Figure 11. Selected analysis regions (blue mesh elements) on the Case 1 spacecraft model and comparison of AO flux results for different surface orientations (+x, -y, +z planes)*

Figure 12 presents the AO flux and cumulative fluence calculated at the selected representative mesh points on each surface. Figure 12(a) shows the temporal variation of the AO flux over the one-year simulation period. The flux exhibits both short-period oscillations and longer-term variations. The short-period oscillations arise as the spacecraft repeatedly samples different latitude, longitude, and local solar time regions along its orbit, whereas the longer-term envelope reflects seasonal variations in the thermospheric AO density predicted by the NRLMSISE-00 model. Among the analyzed surfaces, the ram-facing +x surface exhibited the highest average AO flux of $2.2 \times 10^{14}$ $atoms/cm^2/s$, corresponding to an annual cumulative fluence of $6.9 \times 10^{21}$ $atoms/cm^2$. This is attributed to direct exposure to the incoming AO flow and the high relative velocity of the spacecraft.

As shown in Figure 12(b), the cumulative AO fluence increased monotonically with time, while clear differences in the accumulated exposure were observed depending on surface orientation. Table 5 quantitatively summarizes the average AO flux, cumulative fluence, and predicted erosion depth for the major surface directions. The wake-facing -x surface experienced negligible AO exposure because the incoming AO flow was largely blocked by the spacecraft body, whereas the side and zenith/nadir-facing surfaces received lower fluence than the ram-facing surface because of their reduced projected exposure to the incoming AO flow. Compared with the ram-facing (+x) surface, the cumulative AO fluence was only approximately 4-5% on the side panels and about 4% on the zenith(+z) and nadir (-z) surfaces, highlighting the strong geometry-dependent anisotropy of AO exposure

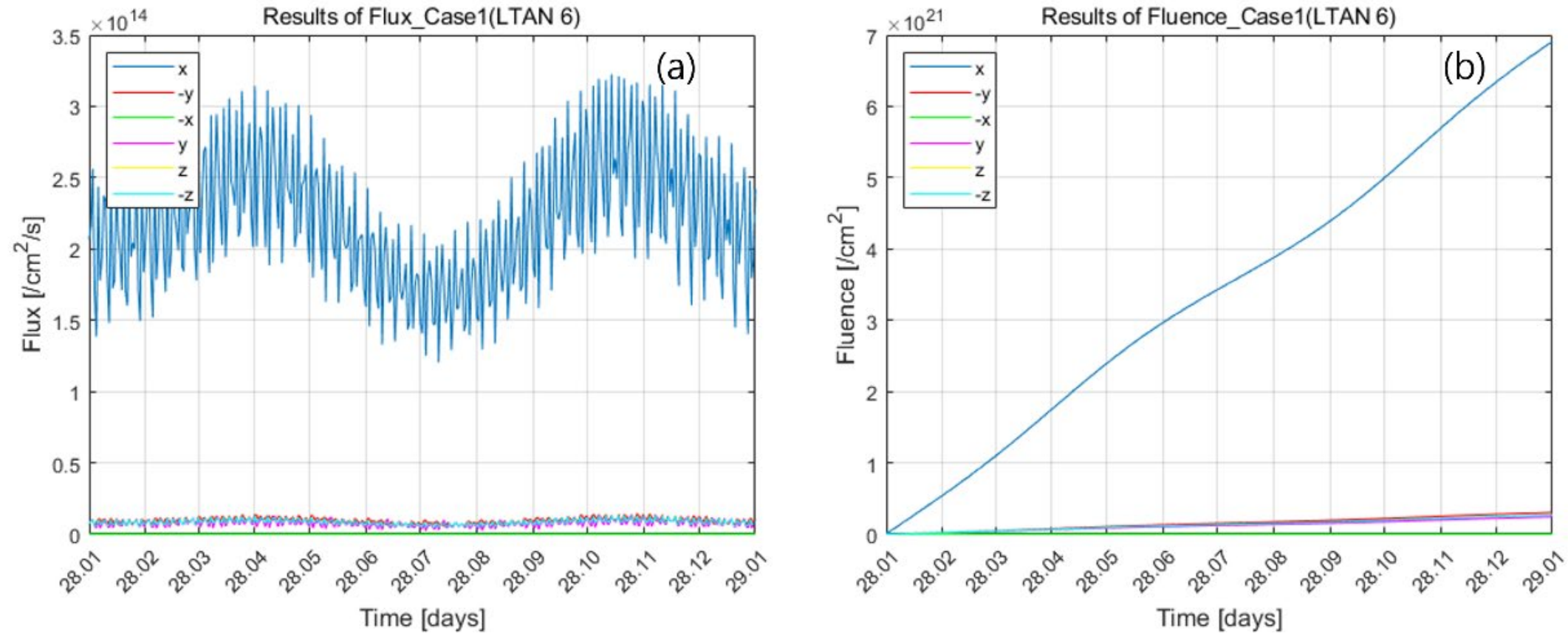

*Figure 12. Surface-specific AO exposure results for the Case 1 spacecraft in the LTAN 06:00 orbit: (a) AO flux and (b) cumulative AO fluence*

| Spacecraft component | Surface | Average flux $(atoms/cm^2/s)$ | Cumulative fluence $(atoms/cm^2)$ | Erosion depth $(\mu m)$ |
|---|---|---|---|---|
| Side panel | +x | $2.2 \times 10^{14}$ | $6.9 \times 10^{21}$ | 0.69 |
| | -x | 0 | 0 | 0 |
| | +y | $7.6 \times 10^{12}$ | $2.4 \times 10^{20}$ | 0.02 |
| | -y | $9.5 \times 10^{12}$ | $3.0 \times 10^{20}$ | 0.03 |
| Solar array | +z | $8.5 \times 10^{12}$ | $2.7 \times 10^{20}$ | 15.1 |
| Antenna | -z | $8.5 \times 10^{12}$ | $2.7 \times 10^{20}$ | 0 |

*Table 5. Average AO flux, cumulative fluence, and predicted erosion depth for the Case 1 spacecraft (LTAN 06:00)*

The erosion depth was estimated using the erosion yield values assigned to each material. Although the +z solar array surface received a lower AO flux than the +x ram-facing MLI surface, it exhibited a relatively large predicted erosion depth because the assigned erosion yield of CFRP was significantly higher than that of the MLI material. This result indicates that AO-induced material degradation is governed not only by the incident AO fluence but also by the material-specific erosion yield.

### 3.1.2. Results for the LTAN 12:00 Case

The same simulation conditions were applied as in the LTAN 06:00 case, except that the LTAN was changed to 12:00. This LTAN was selected to represent a near-noon Earth observation mission scenario commonly used in VLEO. As shown in Figures 13 and 14, the overall AO distribution pattern remained similar to that of the LTAN 06:00 case; however, differences were observed in the magnitude of the AO flux and cumulative fluence on the spacecraft surfaces.

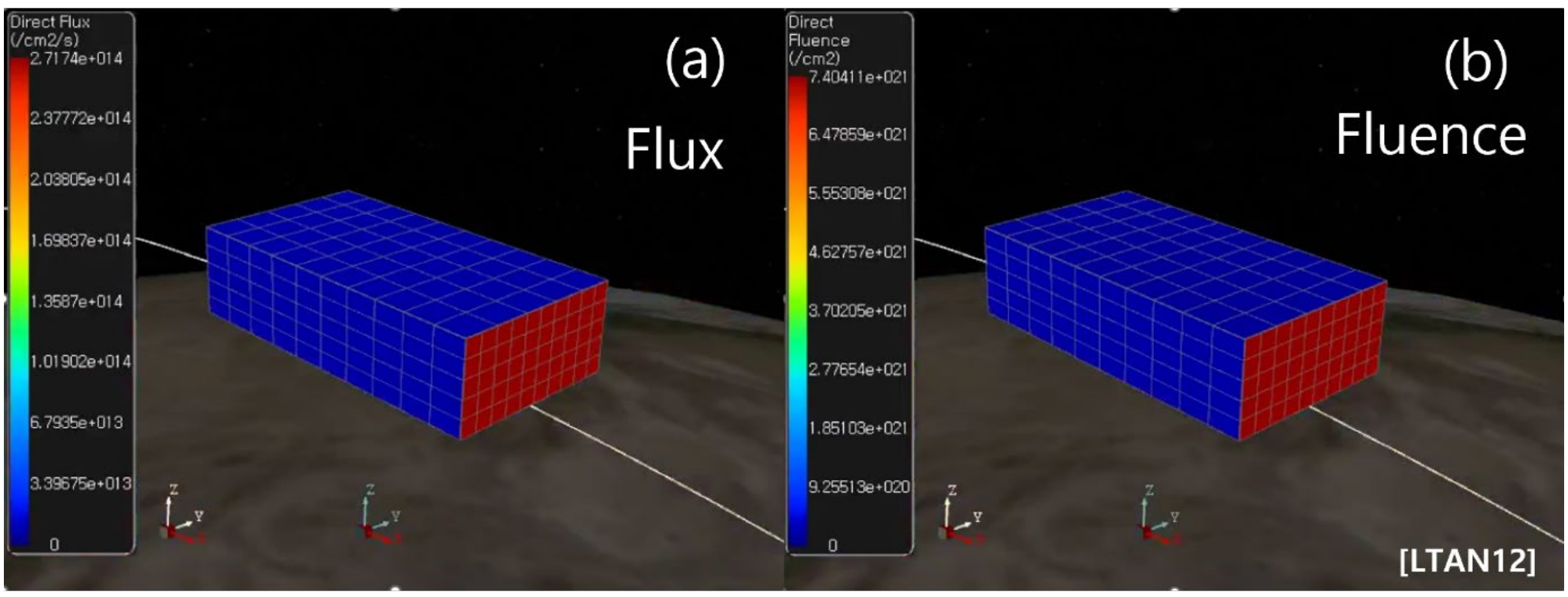

*Figure 13. Simulated AO distribution over the surface of the baseline non-deployable spacecraft model (Case 1) for the LTAN 12:00 orbit: (a) AO flux and (b) cumulative AO fluence*

Consistent with the orbit-averaged results summarized in Table 3, the LTAN 12:00 case showed higher AO density and flux than the LTAN 06:00 case, leading to a higher cumulative fluence on the exposed spacecraft surfaces.

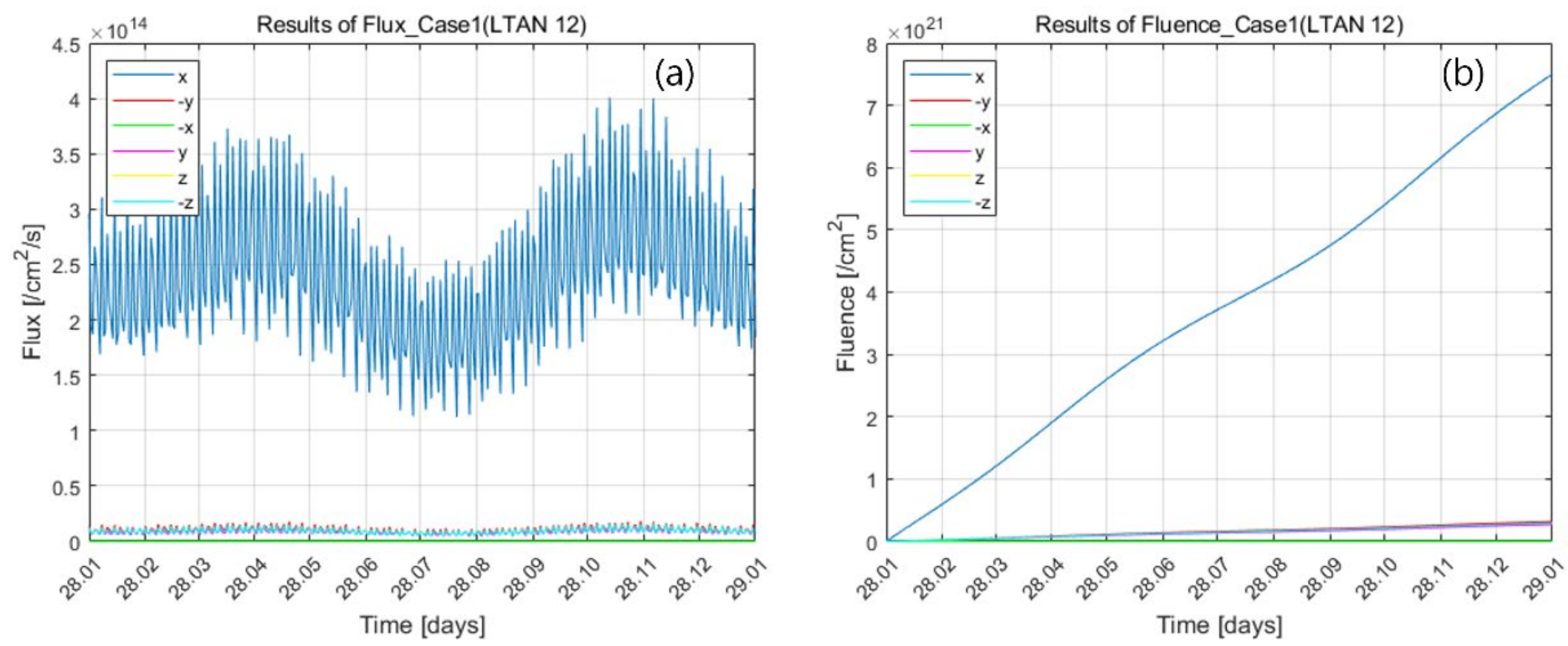

*Figure 14. Surface-specific AO exposure results for the Case 1 spacecraft in the LTAN 12:00 orbit: (a) AO flux and (b) cumulative AO fluence*

| Spacecraft component | Surface | Average flux $(atoms/cm^2/s)$ | Cumulative fluence $(atoms/cm^2)$ | Erosion depth $(\mu m)$ |
|---|---|---|---|---|
| Side panel | +x | $2.4 \times 10^{14}$ | $7.5 \times 10^{21}$ | 0.75 |
| | -x | 0 | 0 | 0 |
| | +y | $8.5 \times 10^{12}$ | $2.7 \times 10^{20}$ | 0.03 |
| | -y | $1.0 \times 10^{13}$ | $3.2 \times 10^{20}$ | 0.03 |
| Solar array | +z | $9.2 \times 10^{12}$ | $2.9 \times 10^{20}$ | 16.2 |
| Antenna | -z | $9.2 \times 10^{12}$ | $2.9 \times 10^{20}$ | 0 |

*Table 6. Average AO flux, cumulative fluence, and predicted erosion depth for the Case 1 spacecraft (LTAN 12:00)*

The results indicate that the cumulative AO fluence for the LTAN 12:00 configuration is consistently higher than that for LTAN 06:00 across all surfaces except for -x direction, with the orbit-averaged flux values summarized in Table 3 showing an increase of approximately 8-10%, which is consistent with the differences observed in Tables 5 and 6 for spacecraft-specific surfaces. While Table 3 provides a scalar, orbit-averaged perspective, the surface-resolved analyses reveal significant directional variations in fluence: the ram-facing +x surface consistently receives the highest cumulative fluence, whereas the wake-facing -x surface exhibits cumulative fluence that is 0% of the ram-facing value and the side panel exhibits approximately 3-4% of the ram-facing value, highlighting the strong anisotropy induced by spacecraft geometry. These findings emphasize that cumulative AO exposure cannot be accurately represented by a single average value but must instead account for the orientation and local geometry of each spacecraft surface. Accordingly, AO resistance and protective coating strategies should be applied selectively based on surface-specific fluence, and mission designers should consider LTAN-dependent differences in AO exposure, which can vary by approximately 8-10% between the 06:00 and 12:00 configurations. This geometry- and orientation-resolved approach enables proactive AO impact assessment during the design phase, allowing for optimized material selection, surface treatment, and configuration planning to ensure the longevity and reliability of VLEO spacecraft.

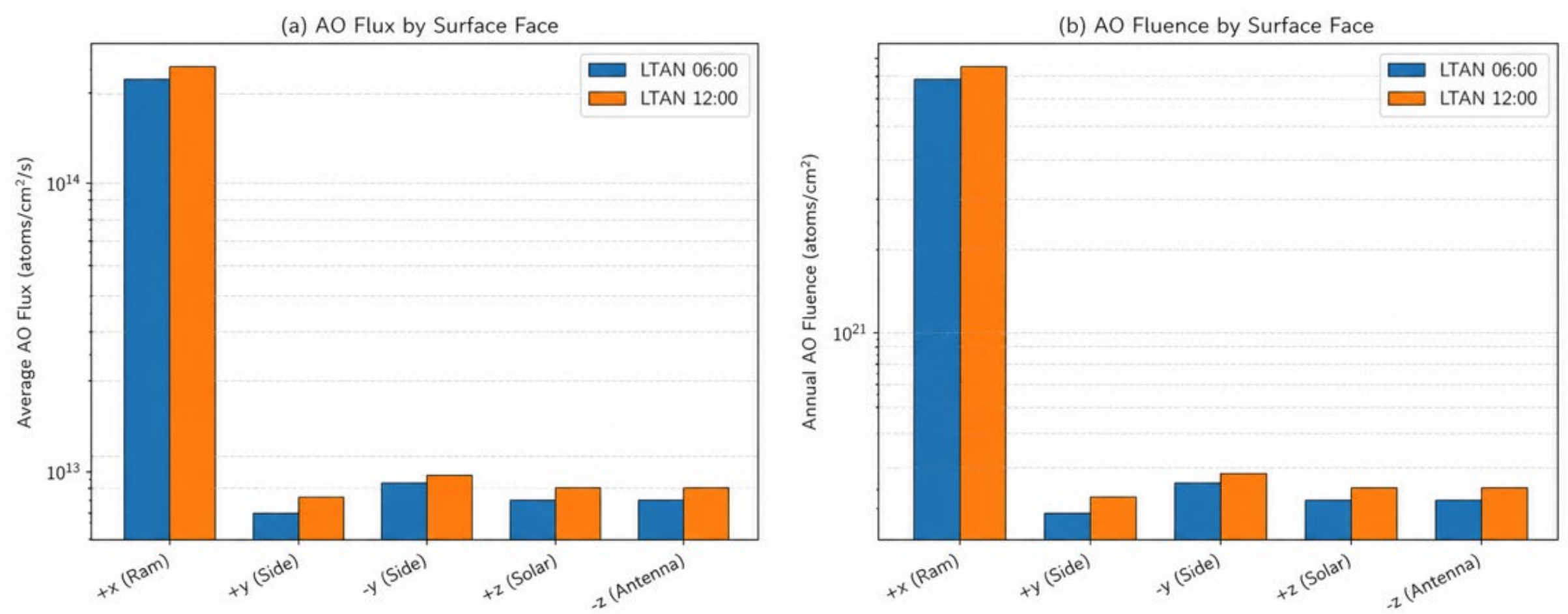

*Figure 15. Comparison of average AO flux and annual fluence by surface face for Case 1 under LTAN 06:00 and LTAN 12:00 conditions (a) AO flux, (b) cumulative AO fluence*

Figure 15 compares the average AO flux and cumulative AO fluence for the major spacecraft surfaces under the LTAN 06:00 and LTAN 12:00 conditions. Consistent with the orbit-averaged AO environment summarized in Table 3, slightly higher flux and fluence values were observed for all exposed surfaces in the LTAN 12:00 case. The ram-facing +x surface remained the dominant AO exposure region under both LTAN configurations, while the wake-facing -x surface experienced negligible AO exposure.

Figure 16 compares the predicted annual erosion depth for the major spacecraft surfaces under the LTAN 06:00 and LTAN 12:00 conditions. The overall erosion trend follows the cumulative AO fluence distribution, with the LTAN 12:00 case producing slightly greater erosion than the LTAN 06:00 case. The largest erosion depth was observed on the +z solar-array surface because of the relatively high erosion yield of CFRP, despite its lower AO flux compared with the ram-facing +x surface. In contrast, the aluminum antenna surface exhibited negligible erosion due to its assumed erosion yield of zero. The predicted erosion depths provide a quantitative estimate of mission-lifetime material degradation and can serve as engineering design references for spacecraft configuration, material selection, and AO mitigation strategies in VLEO missions.

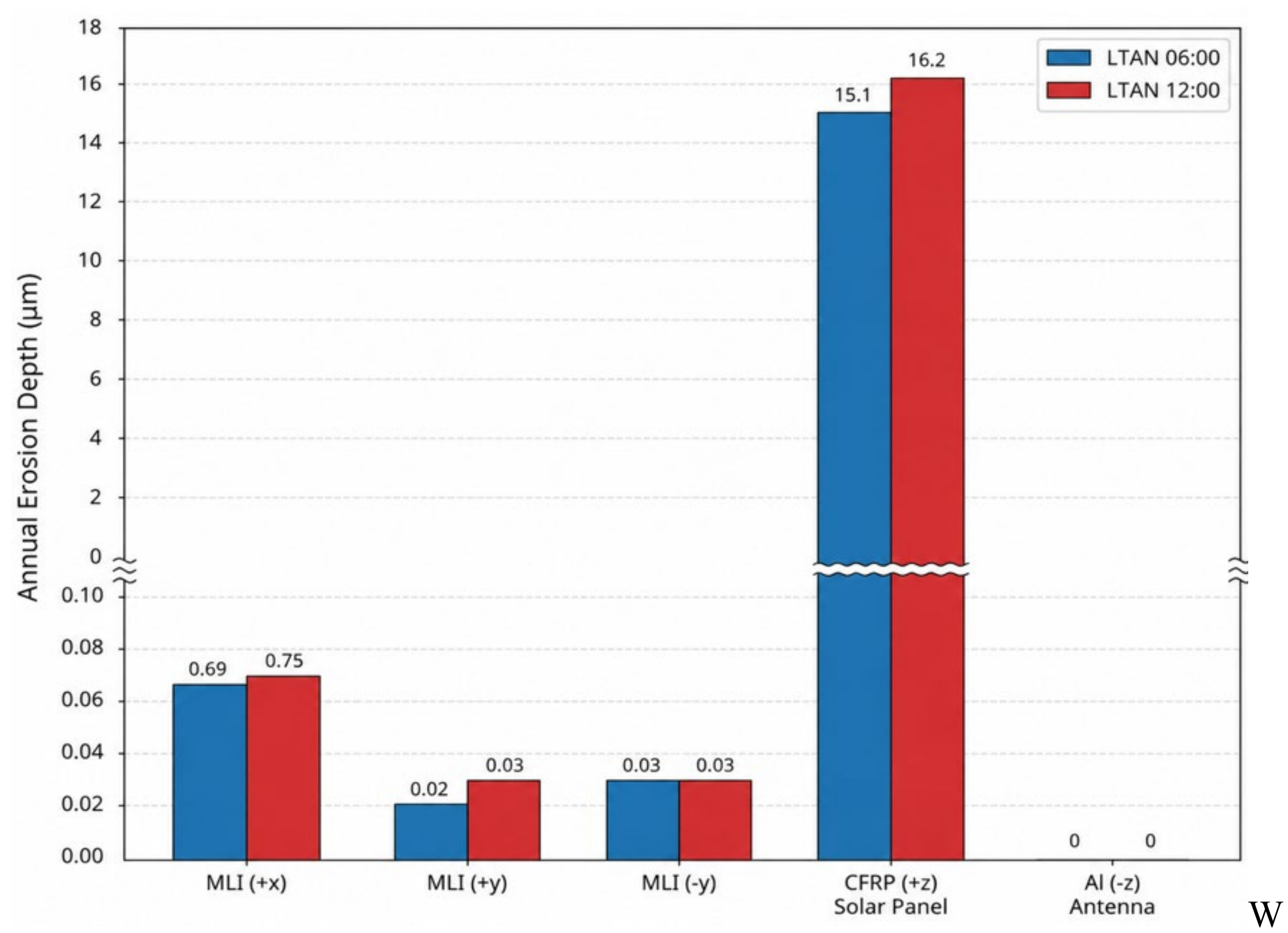

*Figure 16. Comparison of predicted annual erosion depth for Case 1 under LTAN 06:00 and LTAN 12:00 conditions*

| MISSE 8 Orientation | Sample ID | Holder Style | MISSE 8 Fluence $(atoms/cm^2)$ | Relative to Ram Fluence |
|---|---|---|---|---|
| Ram | M8-R1 | Beveled Tray | $4.62 \times 10^{21}$ | 100% |
| Wake | M8-W1 | Beveled Tray | $8.80 \times 10^{19}$ | 1.9% |
| Zenith | M8-Z1B | Beveled Tray | $4.04 \times 10^{19}$ | 0.87% |
| Zenith | M8-1 | Thin Al Foil (Taped) | $1.96 \times 10^{20}$ | 4.24% |

*Table 7. Measured atomic oxygen fluence for MISSE 8 Kapton H exposed in ram, wake, and zenith orientations, normalized to the ram-facing fluence [27]*

The Table 7 provides an important reference for evaluating the geometry-dependent AO fluence distributions predicted by ATOMOX. The holder configuration used in MISSE 8 should first be considered when interpreting the measured fluence values. The beveled tray configuration places the sample within an aluminum holder featuring raised edges, whereas the taped sample configuration mounts the specimen directly onto a thin aluminum foil with minimal surrounding obstruction.

For the zenith-facing exposure, the MISSE 8 taped-sample configuration exhibited a fluence corresponding to approximately 4.24% of the ram-facing fluence. This value is remarkably consistent with the ATOMOX predictions, which indicate that the zenith-facing surfaces receive approximately 4% of the ram-facing AO fluence. In contrast, the zenith-facing sample mounted within the beveled tray exhibited a significantly lower relative fluence of only 0.87% of the ram-facing value. This large discrepancy is likely attributable to AO shielding by the surrounding aluminum holder structure rather than to the actual zenith AO environment. NASA has previously reported that aluminum sample holders can alter local AO exposure conditions, including AO concentration effects near exposed edges under ram-facing conditions [28]. For non-ram-facing surfaces such as zenith orientations, the same holder geometry can instead reduce the effective AO exposure through geometric shadowing and shielding effects. Consequently, the beveled-tray result should not be interpreted as a representative measurement of the intrinsic zenith-facing AO fluence environment. Figure 17 illustrates the zenith-facing beveled tray and taped samples from MISSE 8 and the AO shielding effect produced by a beveled aluminum holder under zenith-facing exposure.

These observations highlight the importance of spacecraft geometry in determining local AO exposure. Even relatively small structural features can substantially modify the AO flux and fluence received by neighboring surfaces. Therefore, AO exposure should not be considered solely as an orbit-level environmental parameter; instead, it should be evaluated at the spacecraft geometry level, where structural configuration and local surface orientation can strongly influence the actual AO environment.

A different trend was observed for the wake-facing surface. The MISSE 8 data indicate a non-zero wake fluence corresponding to approximately 1.9% of the ram-facing exposure, demonstrating that a measurable amount of AO can still reach wake-oriented surfaces in the actual space environment. In contrast, the ATOMOX analysis predicted essentially zero AO fluence on the wake-facing surface. This discrepancy is attributed to the ray-tracing-based AO transport model employed by ATOMOX, in which AO particles travel along deterministic ballistic trajectories and are completely blocked by the spacecraft body. While this assumption is effective for identifying primary AO exposure regions, it appears to underestimate AO transport into wake regions. The comparison therefore suggests that the current ATOMOX methodology may not accurately reproduce wake-side AO exposure and that additional physical mechanisms should be considered to improve wake-region predictions.

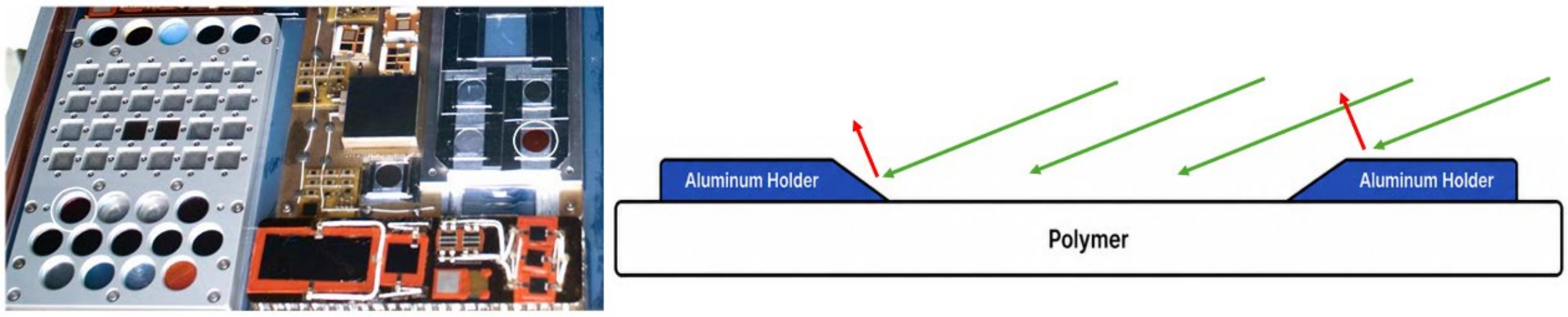

*Figure 17. Close-up image of zenith beveled tray and taped samples on the MISSE 8. The Kapton H samples have been circled. (Left) AO shielding effect of a beveled aluminum holder under zenith-facing exposure. (Right)*

3.2. Influence of Structural Appendages (Case 2)

In a simplified flat-surface spacecraft configuration, atomic oxygen primarily impinges on surfaces exposed to the incoming AO flow, resulting in relatively predictable exposure characteristics. However, real spacecraft typically contain a variety of three-dimensional structural elements that can create partially shadowed regions. As a result, even under identical orbital conditions, the AO exposure can become highly non-uniform depending on the local surface location and geometry. In this section, the influence of such structural features on AO exposure was investigated using a modified spacecraft model in which a wedge-shaped structure was added to the front surface of the baseline non-deployable configuration. Figure 18 illustrates the geometry of the resulting model, designated as Case 2, in which the wedge structure was attached to the +x surface of the baseline spacecraft. The same mesh configuration and orbital conditions used in Case 1 were applied to enable direct comparison between the two cases. The wedge structure was modeled using aluminum (Al). The material properties assigned to each surface and their corresponding erosion yield values are summarized in Table 8.

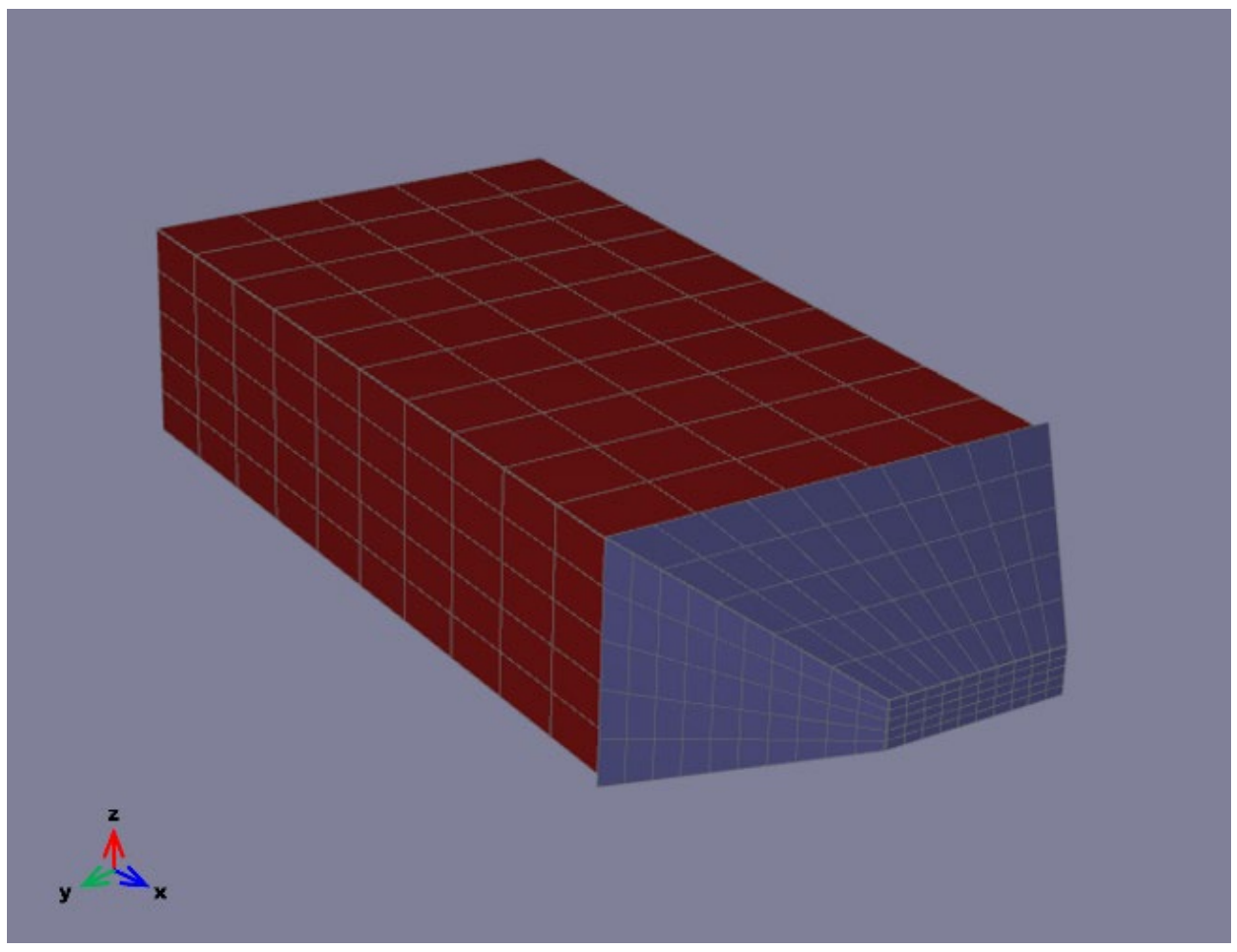

*Figure 18. Geometry of the modified spacecraft model with a wedge-shaped structure attached to the +x surface (Case 2)*

| Spacecraft component | Material | Surface / Orientation | Erosion Yield ($cm^3/atom$) | Ref. |
|---|---|---|---|---|
| Side panel | MLI | +x / Ram | $1.0 \times 10^{-26}$ | [4] |
| | | -x / Wake | | |
| | | +y / Starboard | | |
| | | -y / Port | | |
| Wedge-shaped structure | Al | +x / Ram | 0 | [4] |
| Solar array | CFRP | +z / Zenith | $5.6 \times 10^{-24}$ | [26] |
| Antenna | Al | -z / Nadir | 0 | [4] |

*Table 8. Materials and corresponding erosion yield values assigned to the modified spacecraft model with a wedge-shaped structure attached to the +x surface (Case 2)*

3.2.1 Results for the LTAN 06:00 Case

The geometric configuration of the additional structure and the resulting changes in surface orientation significantly altered the distribution of AO flux and cumulative fluence across the spacecraft surfaces. Analysis of the wedge-structure model (Case 2) revealed noticeable differences in the AO exposure pattern compared with the baseline non-deployable configuration (Case 1). As shown in Figure 19, relatively high AO flux was maintained on the forward-facing surface of the wedge structure because this region was directly exposed to the incoming AO flow along the ram direction. In contrast, reduced AO exposure was observed in the region behind the wedge structure due to the shielding effect introduced by the additional geometry. The cumulative AO fluence distribution exhibited a similar trend, with lower fluence values observed in the shadowed regions located behind the wedge structure.

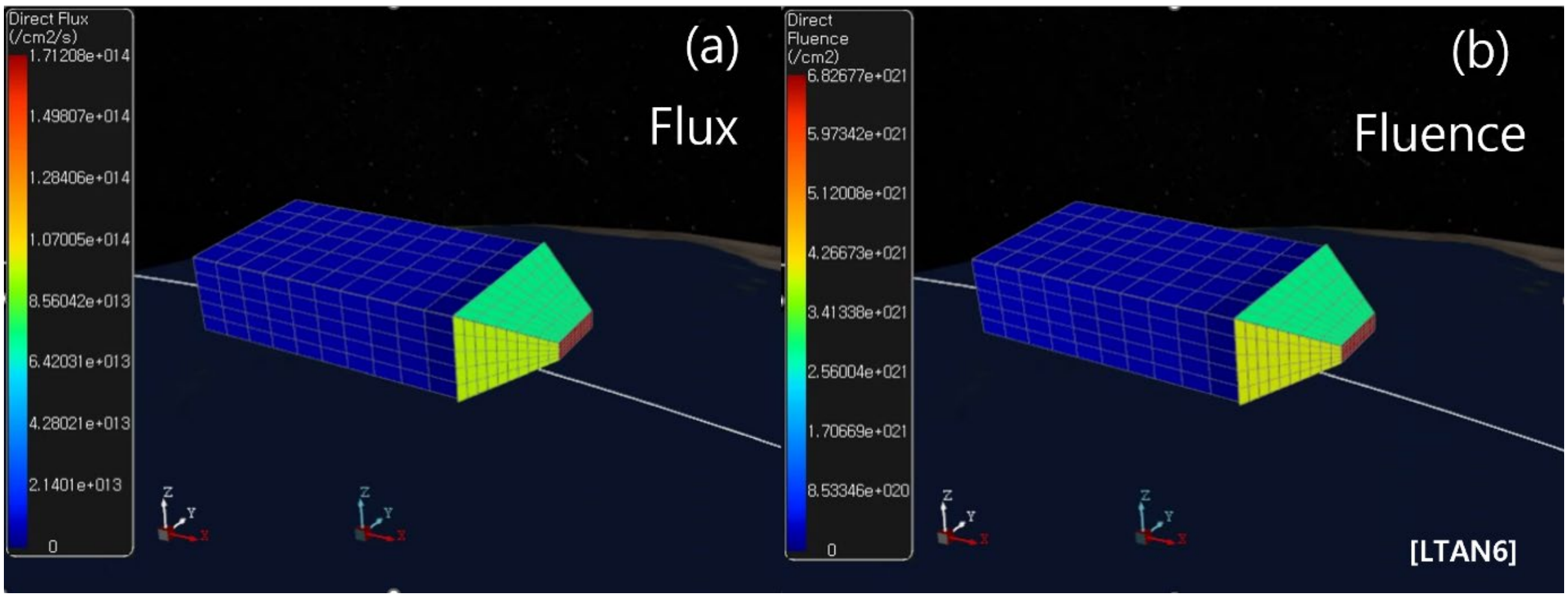

*Figure 19. Simulated AO distribution over the surface of the modified spacecraft model with a wedge-shaped structure attached to the +x surface (Case 2) for the LTAN 06:00 orbit: (a) AO flux and (b) cumulative AO fluence*

To more clearly investigate the AO exposure characteristics of Case 2, the simulation results were analyzed from two different perspectives. As illustrated in Figure 20, the AO distributions were separately evaluated for the wedge structure itself (Figure 20(a)) and the baseline spacecraft body located behind the structure (Figure 20(b)). The results shown in Figure 20(a) indicate that the geometric configuration of the wedge structure produces distinct AO exposure levels depending on surface orientation. The ram-facing surfaces experienced the highest AO flux and fluence, whereas the adjacent inclined surfaces were subjected to relatively lower AO exposure.

As shown in Figure 20(b), the wedge structure generated a shielding effect that reduced AO exposure in the downstream region of the spacecraft body. Comparison of the horizontal surface distribution within the same plane revealed that the AO flux and fluence values were generally similar among neighboring mesh elements. In addition, along the +z surface, the cumulative AO fluence increased progressively toward the -x direction, indicating a spatial exposure gradient across the surface. Approximately one order of magnitude difference was observed between the minimum and maximum cumulative fluence values. In contrast, virtually no AO flux or fluence was detected on the +x surface behind the wedge structure because the incoming AO flow was effectively blocked by the wedge geometry.

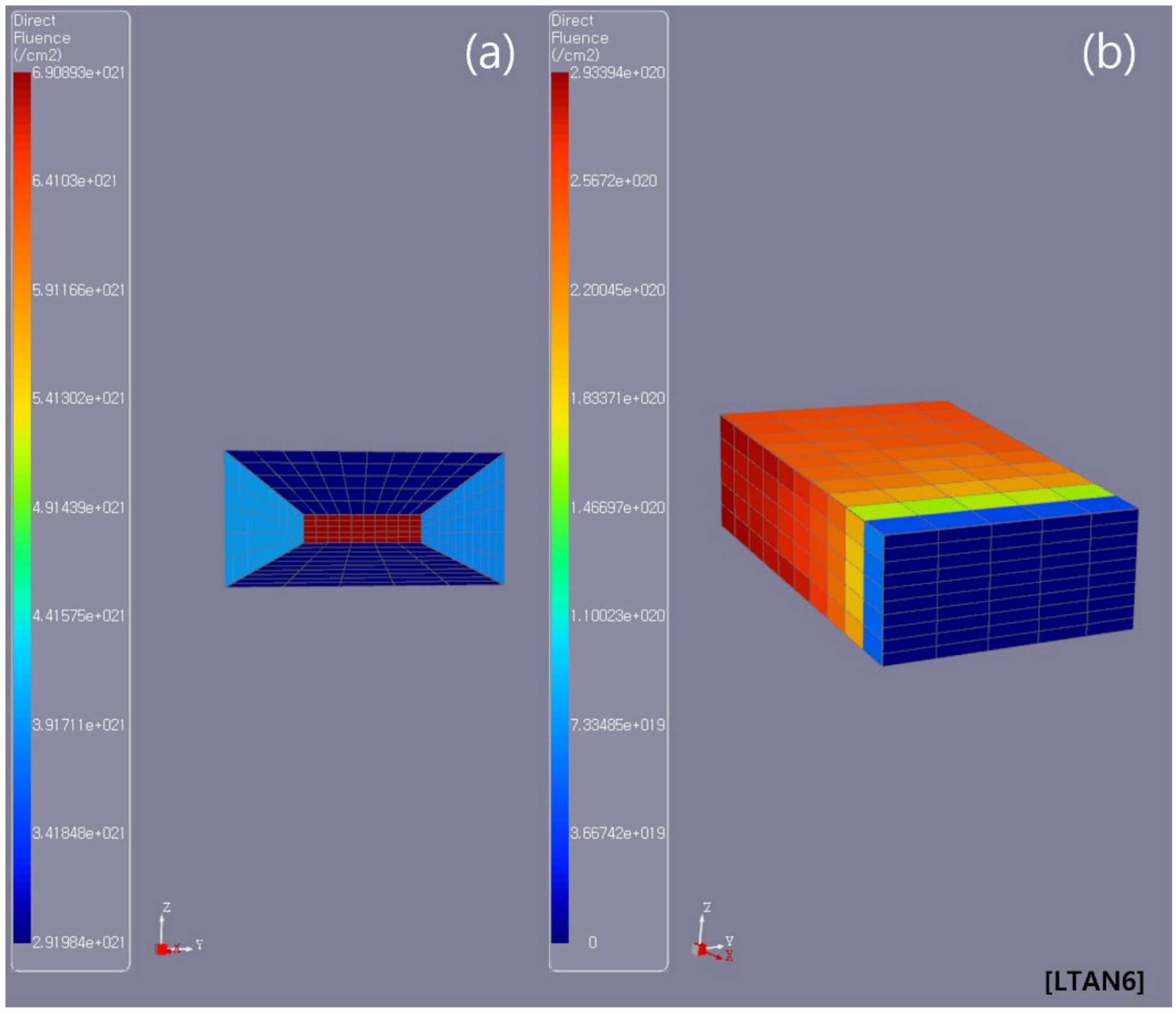

*Figure 20. AO fluence distribution for the Case 2 spacecraft under the LTAN 06:00 condition: (a) wedge structure and (b) baseline spacecraft body*

The material assignments and erosion yield values used in Case 1 were retained for Case 2. As in the baseline configuration, the +z solar-array surface exhibited the largest predicted erosion depth due to the relatively high erosion yield of the CFRP material. Table 9 quantitatively summarizes the average AO flux, cumulative fluence, and predicted erosion depth for the surface directions.

| Spacecraft component | Material | Surface | Average flux $(atoms/cm^2/s)$ | Cumulative fluence $(atoms/cm^2)$ | Erosion depth $(\mu m)$ |
|---|---|---|---|---|---|
| Side panel | MLI | +x | 0 | 0 | 0 |
| | | -x | 0 | 0 | 0 |
| | | +y | $1.1 \times 10^{12}$ $\sim 7.3 \times 10^{12}$ | $3.5 \times 10^{19}$ $\sim 2.3 \times 10^{20}$ | 0.004~0.02 |
| | | -y | $1.7 \times 10^{12}$ $\sim 9.2 \times 10^{12}$ | $5.3 \times 10^{19}$ $\sim 2.9 \times 10^{20}$ | 0.005~0.03 |
| Solar array | CFRP | +z | $1.5 \times 10^{12}$ $\sim 7.9 \times 10^{12}$ | $4.9 \times 10^{19}$ $\sim 2.5 \times 10^{20}$ | 2.7~14 |
| Antenna | Al | -z | $1.4 \times 10^{12}$ $\sim 7.9 \times 10^{12}$ | $4.5 \times 10^{19}$ $\sim 2.5 \times 10^{20}$ | 0 |
| Wedge-shaped structure | Al | +x | $2.2 \times 10^{14}$, $9.2 \times 10^{13}$, $1.3 \times 10^{14}$ | $6.9 \times 10^{21}$, $2.9 \times 10^{21}$, $4.0 \times 10^{21}$ | 0 |

*Table 9. Annual AO flux, cumulative fluence, and predicted erosion depth for Case 2 under the LTAN 06:00 condition*

3.2.2 Results for the LTAN 12:00 Case

This section presents the AO environment analysis results obtained for Case 2 under the LTAN 12:00 condition. The simulation results Figure 21 indicate that the LTAN 12:00 case generally exhibited higher AO flux and cumulative fluence than the LTAN 06:00 case. This trend is consistent with the observations made for Case 1 and reflects the differences in the orbit-averaged AO environment summarized in Table 3. Comparison of the surface-specific distributions further showed that the -y surface experienced relatively higher AO flux than the +z surface. From an erosion perspective, the increase in AO flux and cumulative fluence was directly reflected in the predicted erosion depth. As a result, greater material erosion was predicted under the LTAN 12:00 condition than under the LTAN 06:00 condition. These findings indicate that LTAN can significantly influence the AO exposure environment experienced by spacecraft surfaces and should therefore be considered in spacecraft mission design and operational planning. Table 10 quantitatively summarizes the average AO flux, cumulative fluence, and predicted erosion depth for the surface directions.

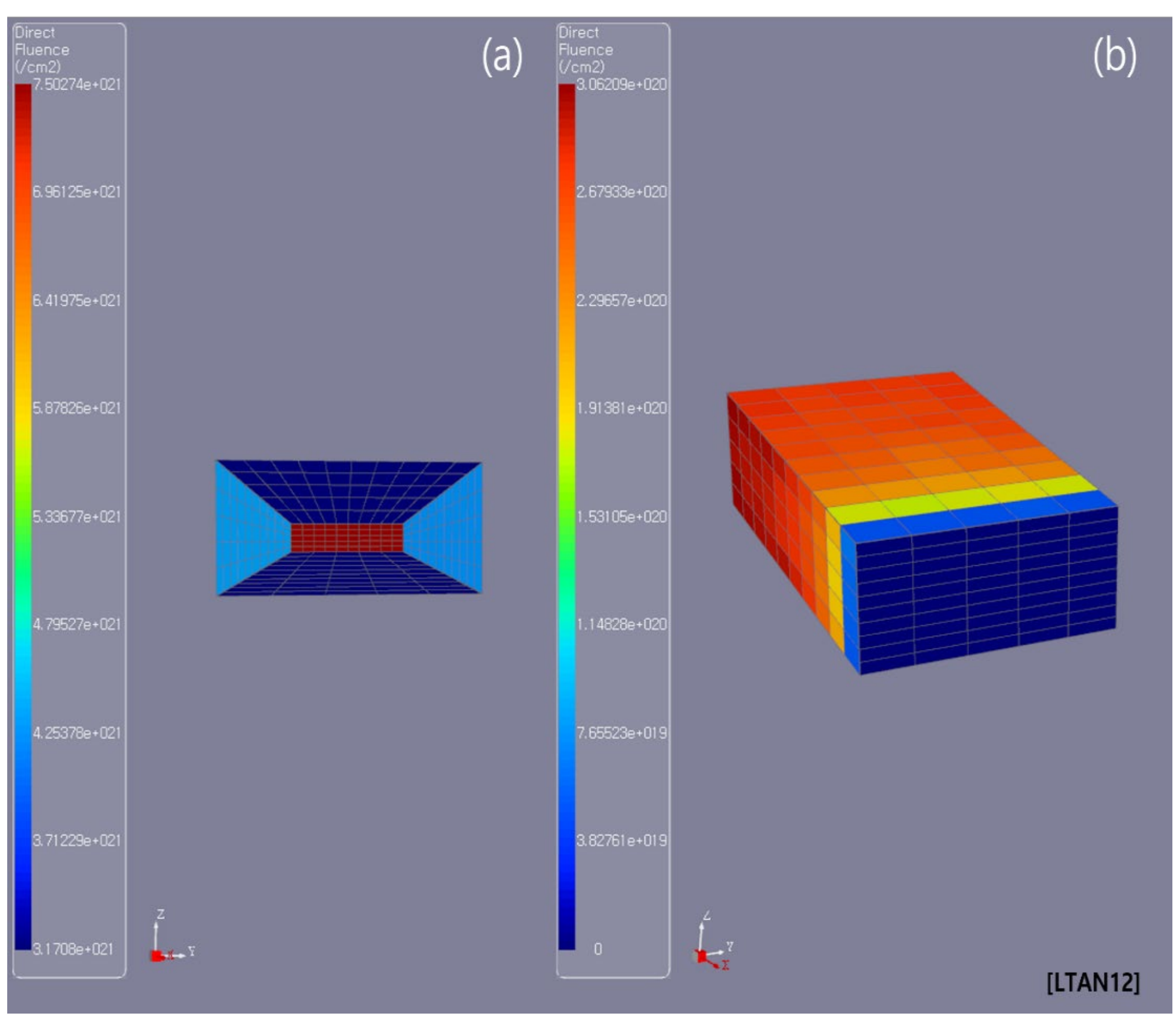

*Figure 21. AO fluence distribution for the Case 2 spacecraft under the LTAN 12:00 condition: (a) wedge structure and (b) baseline spacecraft body*

| Spacecraft component | Material | Surface | Average flux $(atoms/cm^2/s)$ | Cumulative fluence $(atoms/cm^2)$ | Erosion depth ($\mu m$) |
|---|---|---|---|---|---|
| Side panel | MLI | +x | 0 | 0 | 0 |
| | | -x | 0 | 0 | 0 |
| | | +y | $1.2 \times 10^{12}$ ~$7.9 \times 10^{12}$ | $3.9 \times 10^{19}$ ~$2.5 \times 10^{20}$ | 0.004~0.025 |
| | | -y | $1.7 \times 10^{12}$ ~$9.5 \times 10^{12}$ | $5.3 \times 10^{19}$ ~$3.0 \times 10^{20}$ | 0.005~0.03 |
| Solar array | CFRP | +z | $1.5 \times 10^{12}$ ~$8.5 \times 10^{12}$ | $4.9 \times 10^{19}$ ~$2.7 \times 10^{20}$ | 2.7~15 |
| Antenna | Al | -z | $1.5 \times 10^{12}$ ~$8.5 \times 10^{12}$ | $4.6 \times 10^{19}$ ~$2.7 \times 10^{20}$ | 0 |
| Wedge-shaped structure | Al | +x | $2.2 \times 10^{14}$, $9.2 \times 10^{13}$, $1.3 \times 10^{14}$ | $7.5 \times 10^{21}$, $3.1 \times 10^{21}$, $4.2 \times 10^{21}$ | 0 |

*Table 10. Annual AO flux, cumulative fluence, and predicted erosion depth for Case 2 under the LTAN 12:00 condition*

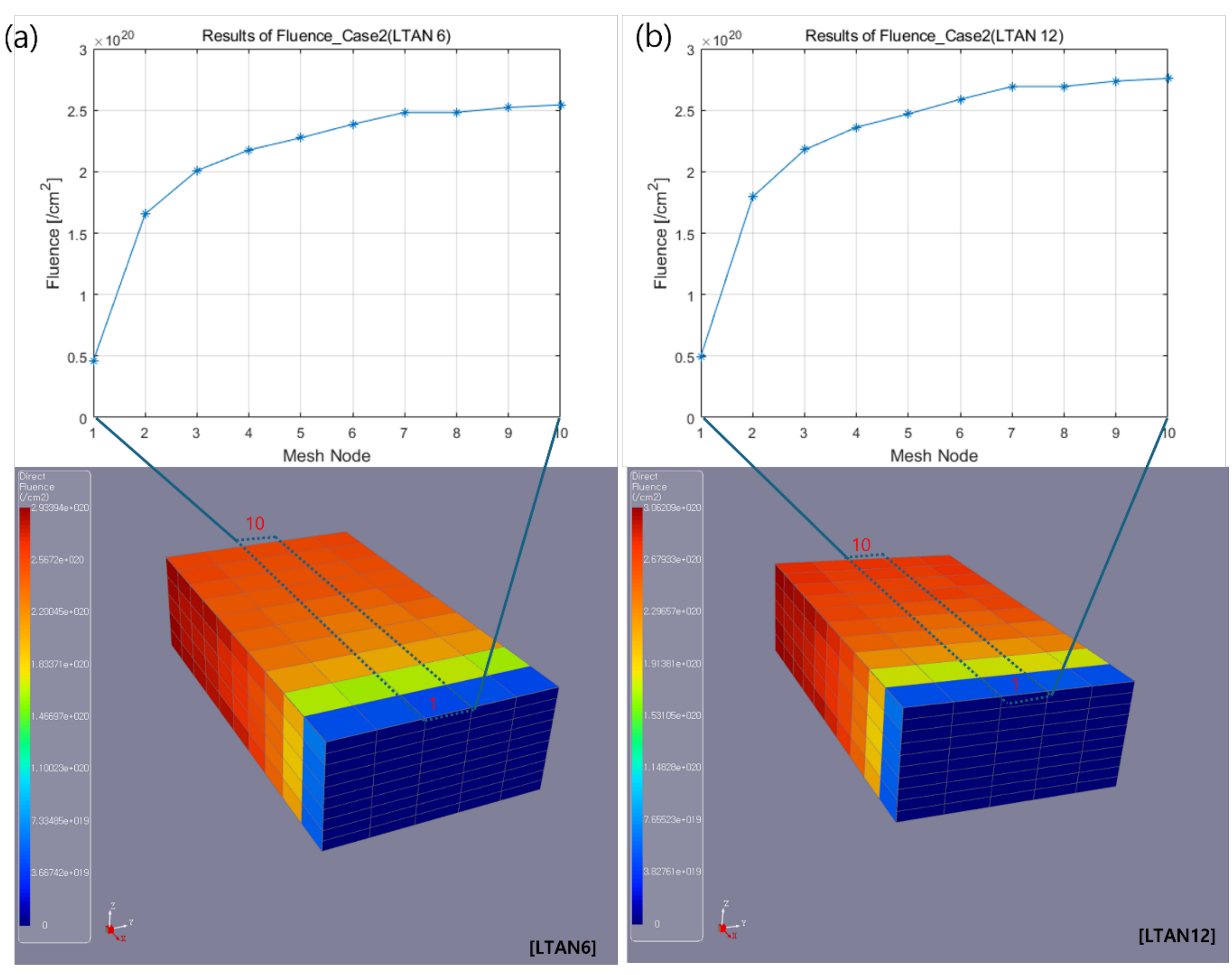

*Figure 22. Comparison of cumulative AO fluence distributions on the baseline spacecraft body of Case 2: (a) LTAN 06:00 and (b) LTAN 12:00*

| Mesh Node | Cumulative fluence (LTAN 06 :00) $(atoms/cm^2)$ $(\times 10^{20})$ | Cumulative fluence (LTAN 12 :00) $(atoms/cm^2)$ $(\times 10^{20})$ |
|---|---|---|
| 1 | 2.687794 | 2.913776 |
| 2 | 2.686357 | 2.912340 |
| 3 | 2.689587 | 2.915394 |
| 4 | 2.686457 | 2.912372 |
| 5 | 2.688022 | 2.913885 |
| 6 | 2.687239 | 2.913128 |
| 7 | 2.687239 | 2.913128 |
| 8 | 2.689587 | 2.915398 |
| 9 | 2.686457 | 2.912372 |

| 10 | 2.684892 | 2.910859 |
| --- | --- | --- |

*Table 11. Comparison of cumulative AO fluence distributions on the baseline spacecraft body of Case 1*

Figure 22 compares the cumulative AO fluence distributions obtained under identical geometric and orbital conditions. The upper plots show the cumulative fluence values extracted at the selected mesh nodes, while the lower color maps visualize the same data as spatial distributions over the spacecraft surface. In both cases, the overall distribution exhibits a similar trend, with the highest cumulative fluence occurring on the downstream surfaces and a sharp reduction in regions affected by geometric shielding. The comparison demonstrates that local geometric features can significantly modify AO exposure, even under identical orbital conditions, by generating shadowed regions and altering the cumulative fluence distribution across the spacecraft surfaces. As summarized in Table 11, the corresponding Case 1 configuration, which does not include the wedge structure, exhibited a maximum node-to-node variation in cumulative AO fluence of less than approximately 0.1%. In contrast, the significantly larger spatial variation observed in Case 2 clearly demonstrates the presence of a strong shielding effect induced by the wedge geometry.

### 3.3. Analysis of Antenna Sub-arrays (Cases 3 and 4)

To evaluate the atomic oxygen exposure characteristics of SAR antenna sub-array structures in the VLEO environment, a geometry-based numerical analysis was performed. Two different antenna configurations were considered to investigate the influence of polarization-dependent structural differences. Case 3 was defined as the H-pol configuration, while Case 4 was defined as the V-pol configuration. The corresponding geometries are illustrated in Figure 23.

The simulation conditions were identical to those used in the previous analyses. A 350 km Sun-Synchronous orbit was assumed, and the analysis period was set to one year, from January 1, 2028 to January 1, 2029. Moderate solar activity conditions were assumed by applying F10.7 as 150 and Ap as 15. The spacecraft attitude was fixed with the +x axis aligned with the velocity vector (ram direction) and the +z axis pointing toward Earth, reflecting the operational configuration in which the SAR antenna is continuously Earth-facing.

In this analysis, the effects of atomic oxygen exposure were evaluated primarily through the distributions of AO flux and cumulative AO fluence rather than predicted erosion depth.

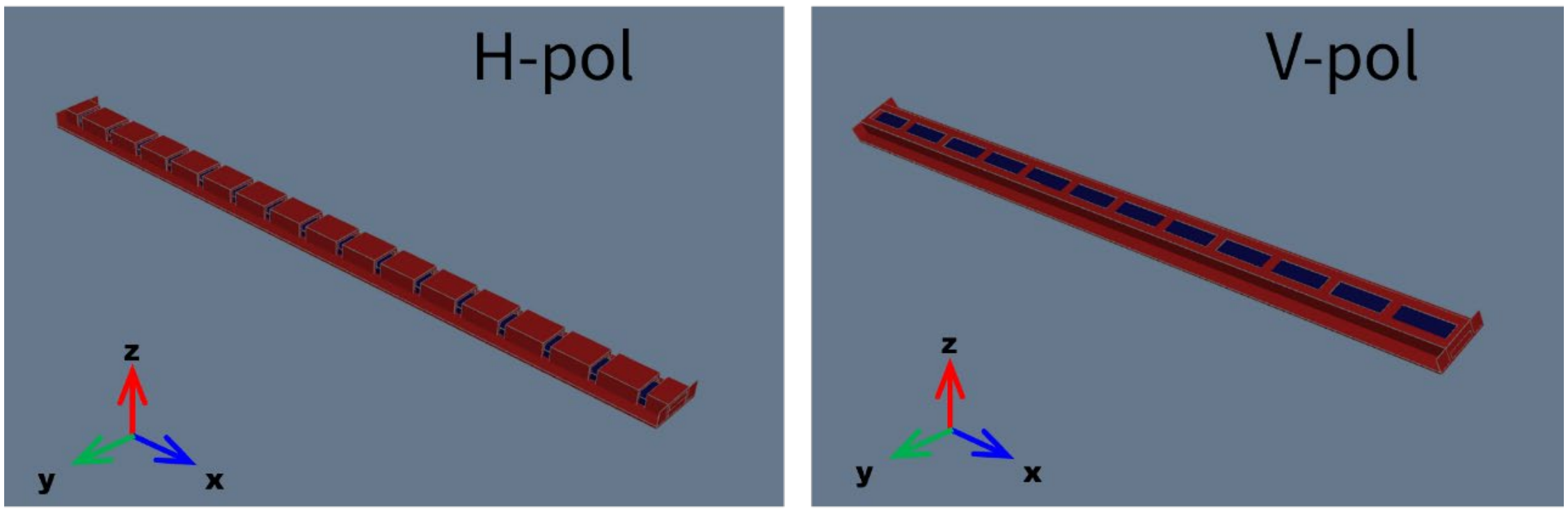

*Figure 23. Geometry configurations of the SAR antenna sub-array models H-pol configuration (Case 3) and V-pol configuration (Case 4)*

### 3.3.1 Results for the H-pol (Case 3) Configuration

To present the analysis results, the evaluation locations defined for the H-pol configuration are shown in Figure 24(a). The Front, Back, and Box Front Side surfaces are oriented toward the ram direction (+x), while the Cover Box and PCB surfaces are oriented toward the Earth-pointing direction (+z). Analysis of the H-pol configuration showed that the highest AO flux and cumulative fluence occurred on the surfaces facing the spacecraft velocity vector (+x). Although the difference in cumulative fluence between the LTAN 06:00 and LTAN 12:00 conditions was approximately $0.5 \times 10^{21}\ atoms/cm^2$, both cases exhibited the greatest AO accumulation on the ram-facing surfaces because of their significantly higher AO flux. In contrast, the surfaces oriented in the +z direction experienced lower AO flux and cumulative fluence due to their reduced projected exposure to the incoming AO flow.

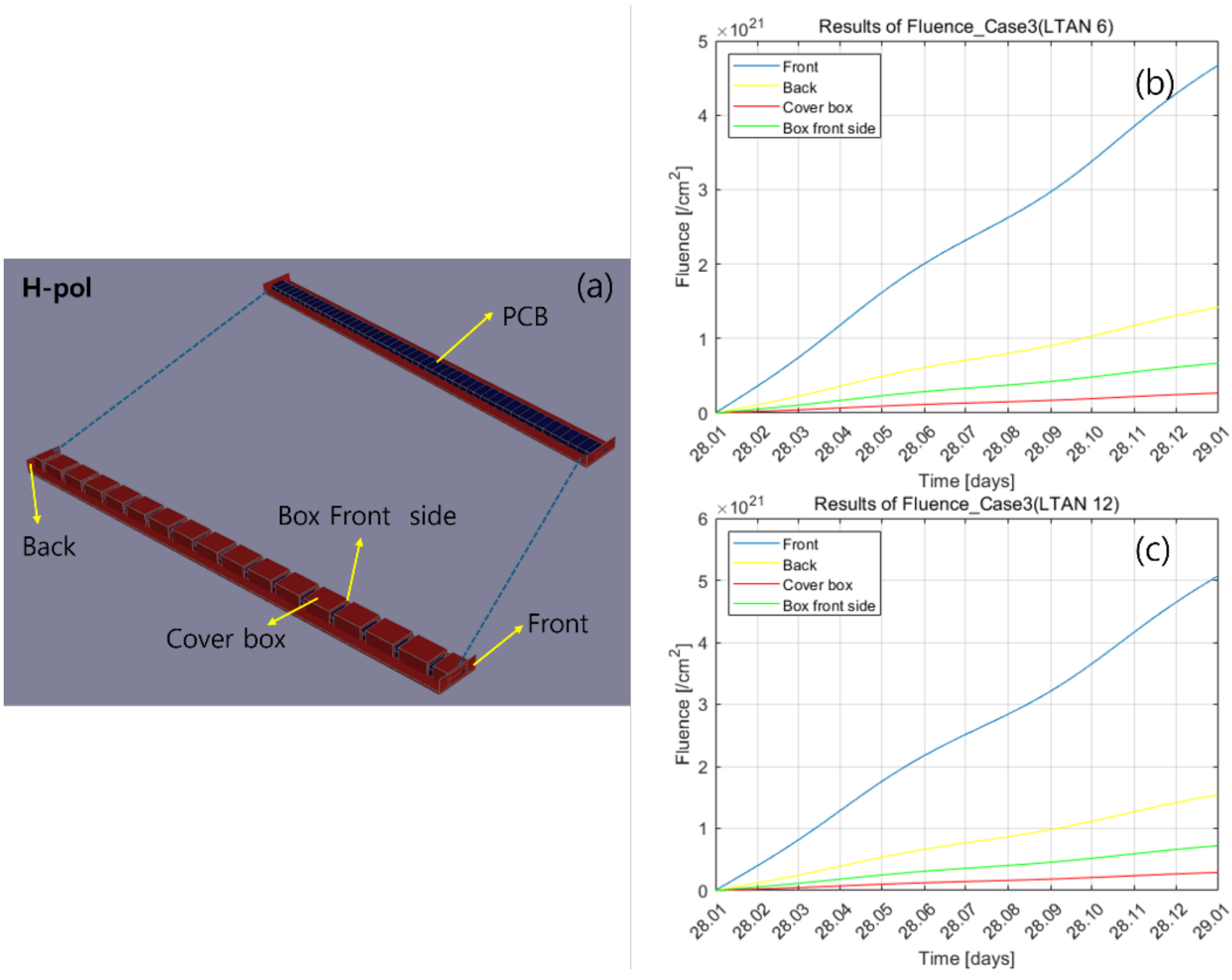

*Figure 24. AO exposure analysis of the H-pol configuration (Case 3): (a) definition of the representative analysis locations, (b) cumulative AO fluence under the LTAN 06:00 condition, and (c) cumulative AO fluence under the LTAN 12:00 condition*

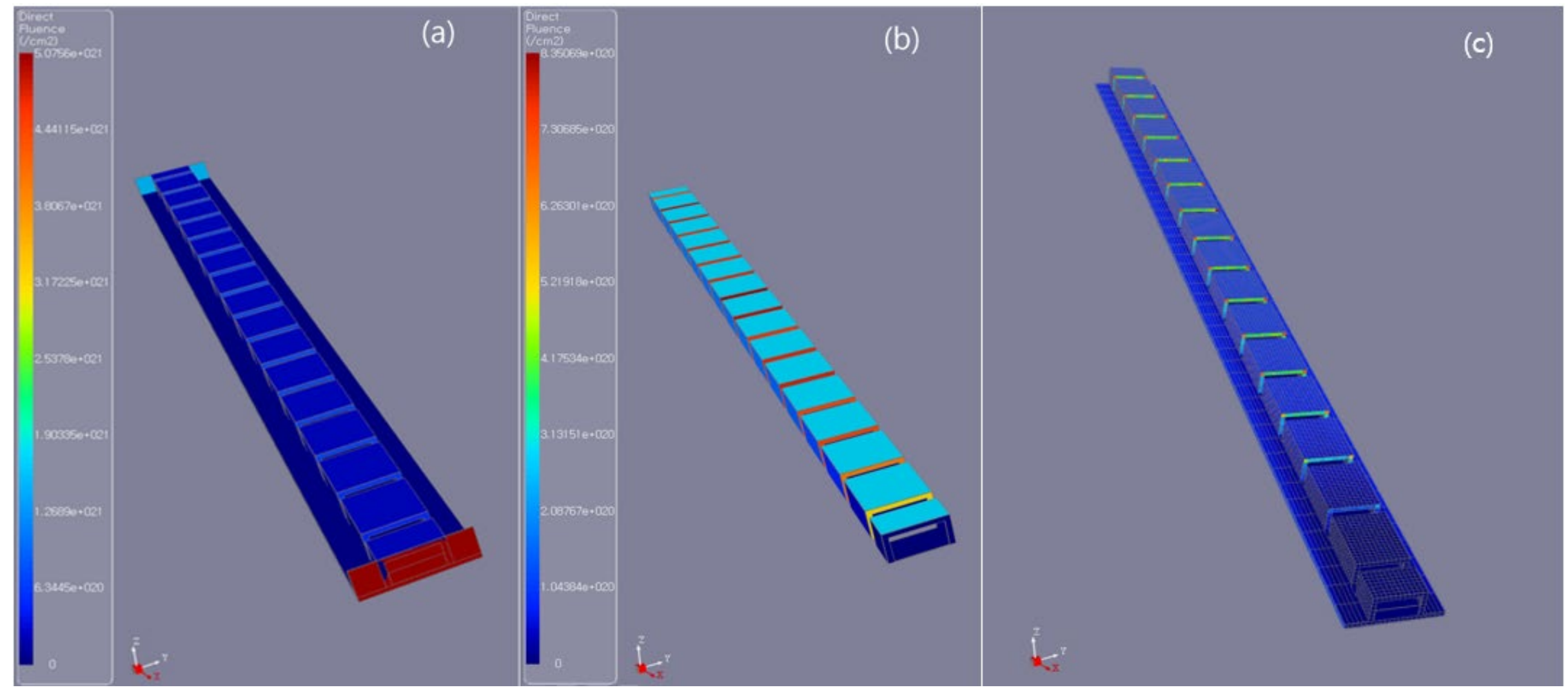

*Figure 25. AO fluence distribution for the H-pol configuration (Case 3) under the LTAN 12:00 conditions: (a) overall structure and (b-c) box region with different mesh size*

Figure 25 presents the cumulative AO fluence distribution for the H-pol configuration under the LTAN 12:00 condition, including the overall structure and a detailed view of the cover-box region. Elevated fluence levels were observed on the forward-facing surfaces directly exposed to the incoming AO flow. The external surfaces of the H-pol structure accumulated an annual AO fluence of approximately $10^{21}$ $atoms/cm^2$ under the LTAN 12:00 orbit. In Figure 25(c), the reduced fluence behind the front structure clearly demonstrates a shielding effect, where the forward-facing element partially blocks the incoming AO flow and suppresses AO accumulation in the downstream region. Table 12 summarizes the average atomic oxygen flux and cumulative AO fluence at different spacecraft surfaces for the H-polarization configuration under the LTAN 12:00 condition.

| Position | Average flux ($atoms/cm^2/s$) | Cumulative fluence ($atoms/cm^2$) |
|---|---|---|
| Front | $2 \times 10^{14}$ | $5 \times 10^{21}$ |
| Back | $3 \times 10^{13}$ | $1 \times 10^{21}$ |
| Cover Box | $6 \times 10^{13}$ | $2 \times 10^{20}$ |
| Box Front Side | $2 \times 10^{13}$ | $7 \times 10^{20}$ |

*Table 12. Average AO flux, cumulative AO fluence for the H-pol configuration under the LTAN 12:00 condition*

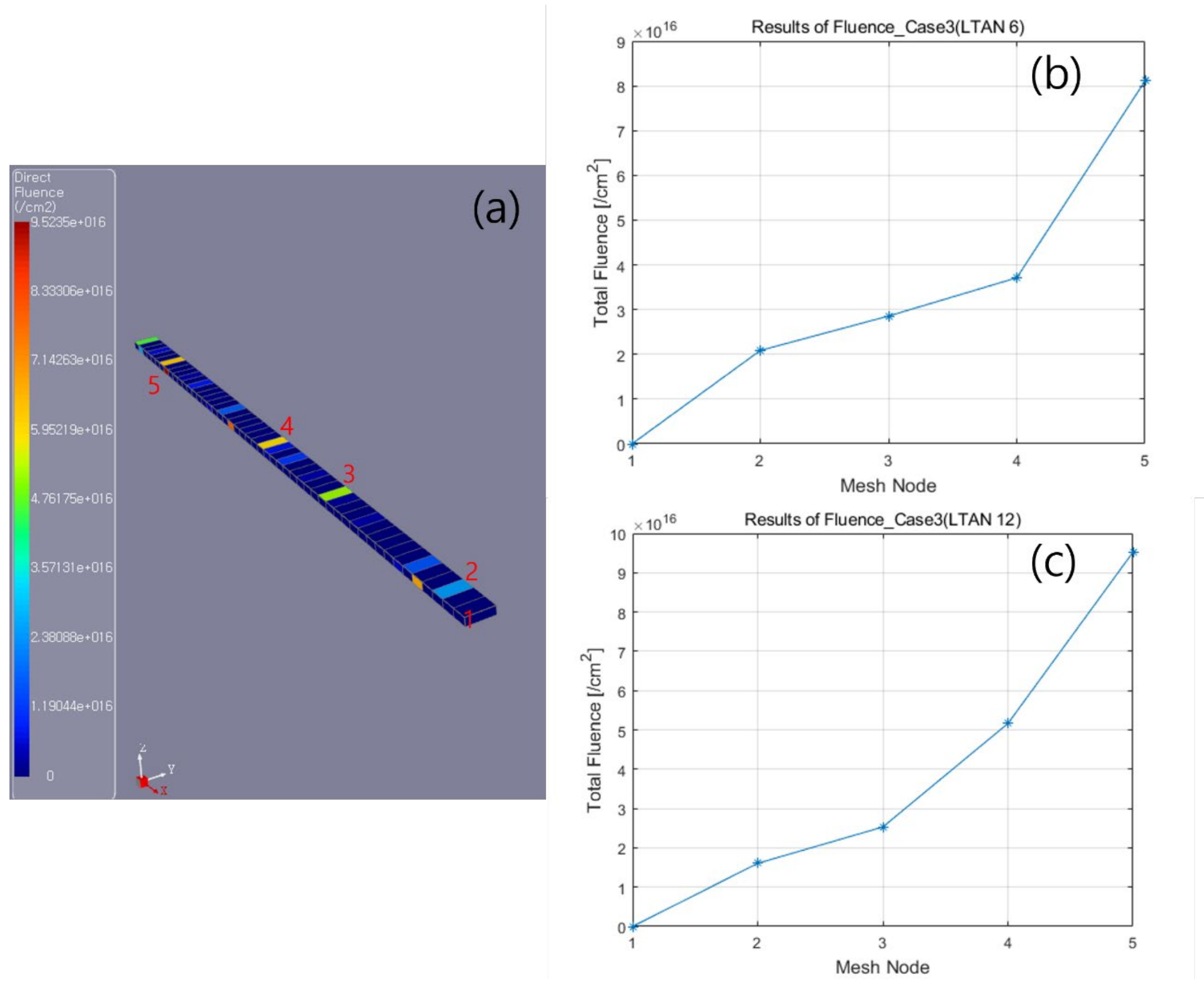

*Figure 26. AO fluence distribution in the PCB region of the H-pol configuration: (a) definition of representative analysis nodes, and (b-c) comparison of AO fluence under LTAN 06:00 and LTAN 12:00 conditions*

The PCB analysis demonstrated that atomic oxygen can enter the internal cavity through openings in the housing structure. Nevertheless, the resulting AO exposure was highly non-uniform across the PCB surface. This non-uniformity is likely associated with the stochastic nature of AO transport, reflection, and penetration paths within the internal cavity during the ray-tracing process. Therefore, rather than interpreting the fluence at each individual mesh node as a deterministic local value, the result should be regarded as evidence that internal components can be exposed to AO when openings exist in the spacecraft structure. As illustrated in Figure 26, the cumulative AO fluence varied significantly with location, ranging from approximately $0$ to $9.5 \times 10^{16}$ $atoms/cm^2$. These results highlight the importance of including internal cavity and opening effects in AO environment assessments for VLEO spacecraft. From an engineering perspective, the maximum predicted internal fluence can be used as a conservative worst-case reference for evaluating material compatibility, coating durability and protection strategies for internal components.

### 3.3.2 Results for the V-pol (Case 4) Configuration

The V-pol configuration was analyzed using the same methodology applied to the H-pol configuration, with representative analysis locations defined as shown in Figure 27. Figure 28 presents the AO fluence distributions for the V-pol configuration under the LTAN 12:00 condition. The Front and Back surfaces correspond to the ram-facing (+x) direction, whereas the Cover Box and PCB regions are oriented toward the Earth-pointing (+z) direction. The resulting AO exposure characteristics were generally consistent with those observed for the H-pol configuration. The highest AO flux and cumulative fluence were observed on the ram-facing surfaces, with the front cover region acting as the dominant exposure area. Compared with the side surfaces of the housing structure, the upper surfaces accumulated relatively higher AO fluence. The maximum cumulative AO fluence was approximately $3 \times 10^{20} atoms/cm^2$. Table 13 summarizes the average atomic oxygen flux and cumulative AO fluence at different spacecraft surfaces for the H-polarization configuration under the LTAN 12:00 condition.

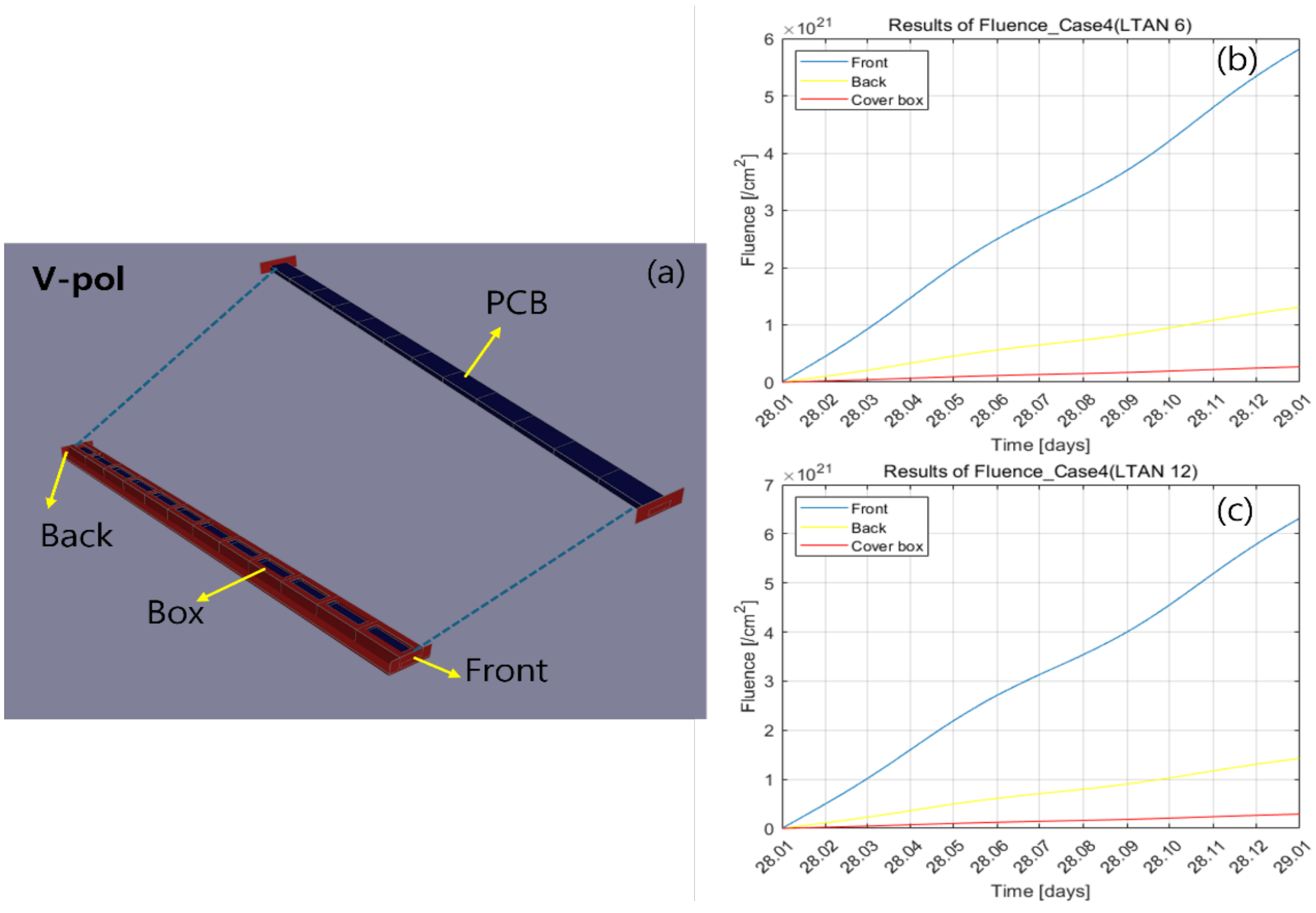

*Figure 27. AO exposure analysis for the V-pol configuration (Case 4): (a) definition of representative surface locations, (b) cumulative AO fluence under the LTAN 06:00 condition, and (c) cumulative AO fluence under the LTAN 12:00 condition*

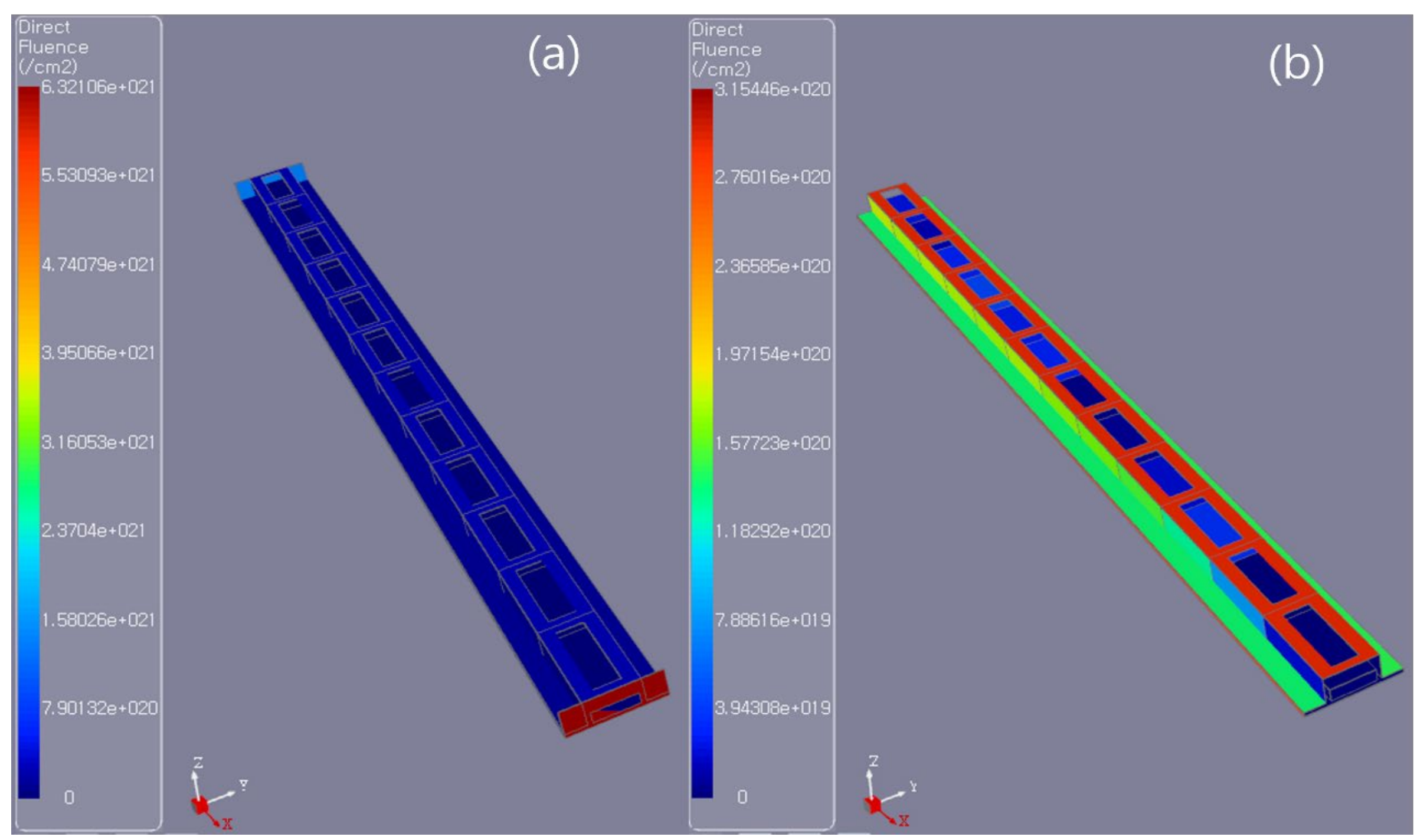

*Figure 28. AO fluence distribution for the V-pol configuration (Case 4) under the LTAN 12:00 conditions: (a) overall structure and (b) box region*

| Position | Average flux ($atoms/cm^2/s$) | Cumulative fluence ($atoms/cm^2$) |
|---|---|---|
| Front | $2 \times 10^{14}$ | $6 \times 10^{21}$ |
| Back | $3 \times 10^{13}$ | $1 \times 10^{21}$ |
| Cover Box | $6 \times 10^{12}$ | $2 \times 10^{20}$ |

*Table 13. Average AO flux, cumulative AO fluence for the V-pol configuration under the LTAN 12:00 condition*

The PCB analysis revealed that AO exposure occurred across most of the PCB surface, with the exception of a small region located at the foremost part of the structure due to shielding effect by Front structure. The average AO flux was approximately $6.3 \times 10^{11} atoms/cm^2/s$, and the annual cumulative AO fluence varied from 0 to approximately $4 \times 10^{19} atoms/cm^2$. In contrast to the external ram-facing surfaces, no distinct spatial trend was observed in the AO distribution across the PCB. This behavior suggests that the PCB exposure is primarily governed by local shielding effects and the geometry-dependent transport of AO through housing openings. The results also indicate that the configuration of enclosure openings can significantly influence the AO environment experienced by internal spacecraft components.

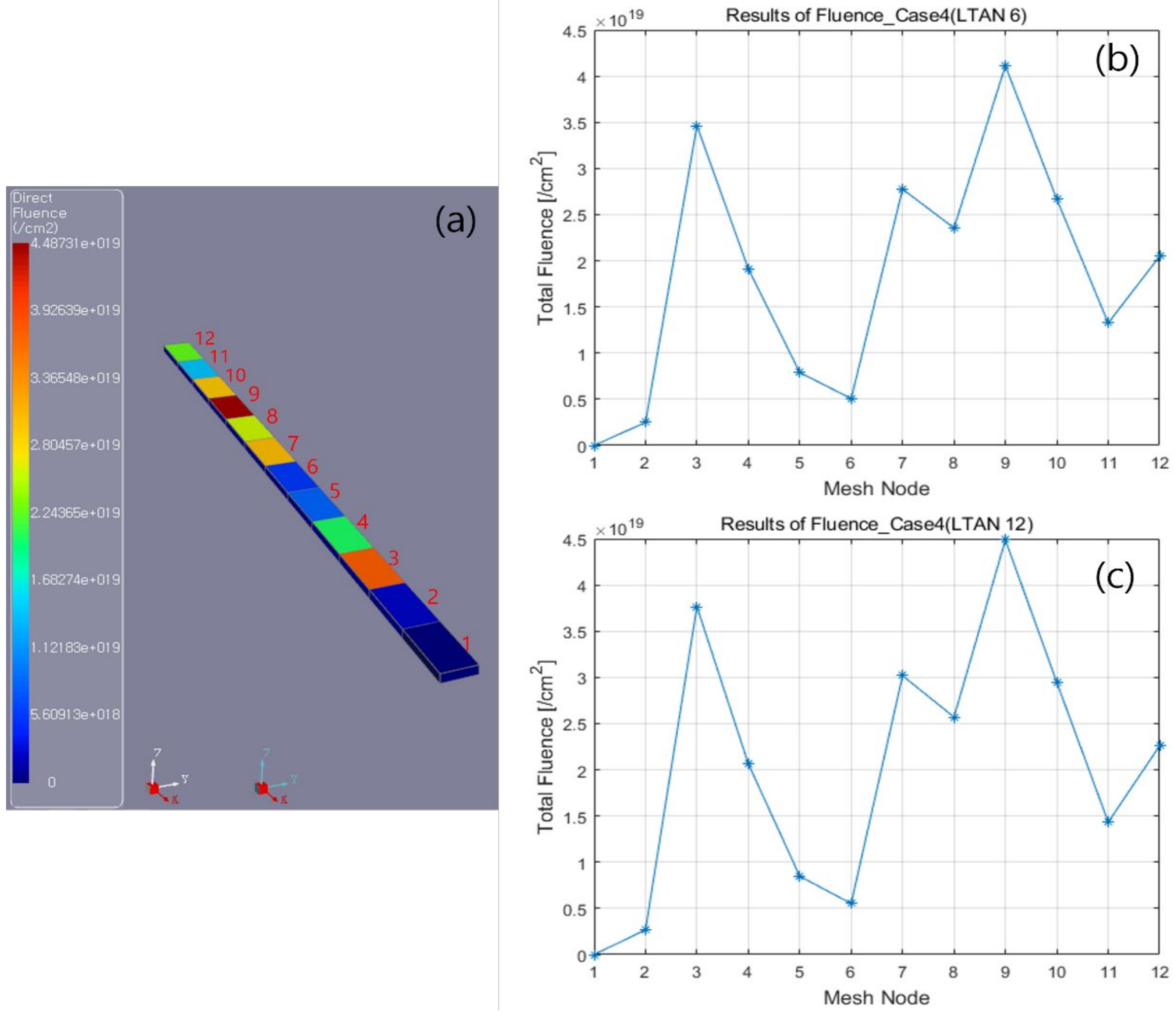

*Figure 29. AO fluence distribution in the PCB region of the V-pol configuration: (a) definition of representative analysis nodes, and (b-c) comparison of AO fluence under LTAN 06:00 and LTAN 12:00 conditions*

A critical finding from this analysis was the penetration of AO into the internal cavities of the antenna structures. As shown in Figure 29, the simulations revealed that AO could enter through openings in the housing and reach the internal Printed Circuit Boards (PCBs). Although the flux levels on the PCBs were significantly lower than on the external ram-facing surfaces, the cumulative fluence over a multi-year mission could still pose a risk to sensitive electronic components or thin protective coatings. The distribution of AO on the PCBs was highly non-uniform, dictated by the internal geometry and the specific pathways available for the gas molecules. This finding aligns with recent studies emphasizing the need to evaluate the internal AO environment in VLEO satellites [29].

## 3.4. Differential AO Exposure on Side Panels under LTAN 12 Conditions

Inspection of the results presented in Tables 5 and 6 revealed a consistent difference in cumulative AO fluence between the +y and -y side panels. Similar asymmetries were also observed in other simulation cases. In Table 6, the difference between the two side panels was approximately 16%, while Table 5 showed a difference of approximately 20%. Considering that the spacecraft dimensions are on the order of a few meters, such differences cannot be attributed to spatial variations in atmospheric density across the spacecraft geometry.

Furthermore, the observed asymmetry persisted under different spacecraft sizes, mission durations, local solar time configurations, and solar and geomagnetic activity conditions, indicating that the effect was systematic rather than numerical noise.

To identify the source of this asymmetry, additional simulations were performed with the HWM07 wind model disabled. Under these conditions, the cumulative AO fluence on the +y and -y side panels became nearly identical, with differences reduced to within approximately 1%. This behavior was consistently observed regardless of spacecraft size, mission duration, local solar time, configurations, and solar and geomagnetic activity conditions. The results therefore indicate that the asymmetry originates from the atmospheric wind model rather than from geometric or numerical effects.

HWM07 is an empirical horizontal wind model that is automatically applied within the ATOMOX framework. The model modifies the effective relative velocity between atmospheric particles and the spacecraft by introducing a horizontal wind component. In ATOMOX, the relative particle velocity is expressed as [23] [25]

$$\vec{v} = \overrightarrow{v_{aero}} + \vec{u} \quad (2)$$

where $\vec{u}$ represents the thermal motion of atmospheric particles and

$$\overrightarrow{v_{aero}} = \overrightarrow{v_{wind}} - \overrightarrow{v_{s/c}}. \quad (3)$$

When the wind model is disabled, the atmosphere is effectively treated as stationary with respect to the spacecraft, and AO particles approach primarily from the ram direction. When the wind model is enabled, the horizontal wind component modifies the effective particle arrival direction. Because ATOMOX evaluates AO transport using a ray-tracing approach, this change in arrival direction alters the ray trajectories and produces measurable differences in cumulative AO fluence on horizontally oriented surfaces such as the ±x and ±y panels.

| LTAN | Mission duration | +y direction Cumulative fluence $(atoms/cm^2)$ | -y direction Cumulative fluence $(atoms/cm^2)$ |
|---|---|---|---|
| LTAN 06:00 | 6 months | $1.334065 \times 10^{20}$ | $1.334233 \times 10^{20}$ |
| | 12 months | $2.687807 \times 10^{20}$ | $2.688165 \times 10^{20}$ |
| LTAN 12:00 | 6 months | $1.442046 \times 10^{20}$ | $1.442363 \times 10^{20}$ |
| | 12 months | $2.913371 \times 10^{20}$ | $2.913827 \times 10^{20}$ |

*Table 14. Comparison of cumulative AO fluence between the +y and -y side panels for different mission durations under LTAN 06:00 and LTAN 12:00 conditions (HWM07 wind model disabled)*

| LTAN | Mission duration | +y direction Cumulative fluence $(atoms/cm^2)$ | -y direction Cumulative fluence $(atoms/cm^2)$ |
|---|---|---|---|
| LTAN 06:00 | 1 month | $1.73 \times 10^{19}$ | $2.03 \times 10^{19}$ |
| | 3 months | $6.27 \times 10^{19}$ | $7.34 \times 10^{19}$ |

| | 6 months | $1.22 \times 10^{20}$ | $1.47 \times 10^{20}$ |
|---|---|---|---|
| | 9 months | $1.77 \times 10^{20}$ | $2.15 \times 10^{20}$ |
| | 12 months | $3.02 \times 10^{20}$ | $2.43 \times 10^{20}$ |
| | 24 months | $4.88 \times 10^{20}$ | $5.93 \times 10^{20}$ |

*Table 15. Comparison of cumulative AO fluence between the +y and -y side panels for different mission durations under LTAN 06:00 condition (HWM07 wind model enabled)*

The results summarized in Table 14 and 15 were generated under the same orbital conditions applied throughout this study. In all cases, the simulation start epoch was set to 1 January 2028. Table 14 and 15 compare the results obtained without and with the HWM07 wind model, respectively. Without the wind model, the difference in AO fluence between the +y and -y surfaces remained below 1%. When the wind model was enabled, the difference increased to approximately 10-20%, depending on the simulation case. In contrast, the $\pm$z surfaces showed only negligible differences because HWM07 represents horizontal atmospheric winds and therefore has little influence on the Earth-pointing and zenith-facing directions. The residual differences observed when the wind model was disabled are considered to represent the numerical uncertainty of the ATOMOX ray-tracing simulation. Since both cases were simulated using 2,000 rays, the resulting uncertainty is estimated to be on the order of less than 1%. These results indicate that the observed asymmetry between the +y and -y surfaces originate primarily from the HWM07 wind model rather than from geometric effects or numerical errors.

## 4. Conclusion

As discussed throughout this report, Very Low Earth Orbit (VLEO) satellites are emerging as a key enabling technology for next-generation space missions. However, the VLEO environment presents a significant engineering challenge in the form of atomic oxygen (AO) exposure, which can lead to material erosion, surface degradation, and potential performance deterioration, making AO one of the primary design constraints for VLEO spacecraft.

In this study, the effects of AO were investigated using a combination of atmospheric environment modeling and geometry-based numerical simulations. Using the NRLMSISE-00 atmospheric model, altitude-dependent AO density and flux were first quantified, followed by spacecraft geometry-resolved analysis with four configurations: a baseline non-deployable shape, a wedge-modified configuration, and two SAR antenna sub-arrays. Two representative Local Time of Ascending Node (LTAN) conditions (06:00 and 12:00) were considered to capture local-time-dependent AO variation.

The results showed that AO flux is strongly concentrated on ram-facing surfaces, while non-ram surfaces received approximately 4% or less of the ram-facing cumulative fluence. Fluence also varied with LTAN, emphasizing the influence of orbital local time. Geometric modifications, such as appendages and antenna assemblies, significantly altered the local AO environment through shielding effects and non-uniform exposure patterns, and atmospheric winds generated measurable side-panel asymmetries. These findings demonstrate that AO durability in VLEO should be treated as a geometry-, material-, and orbit-dependent design problem rather than a scalar environmental load.

The present study highlights the importance of performing quantitative AO environment assessments that account for orbital conditions, spacecraft attitude, and realistic three-dimensional geometry. AO flux and fluence-based durability assessments should be incorporated into material selection and protection strategy development to guide spacecraft design. This geometry-resolved AO risk assessment framework provides actionable guidance for protective coating placement, material selection, and AO mitigation strategies, enabling more reliable and durable VLEO spacecraft.

## Acknowledgements

This work was supported by a 2025 grant from Hanwha Systems and by the National Research Foundation of Korea (NRF) grant funded by the Korea government (MSIT) (No. RS-2025-02213804).